\documentclass[prd,11pt,tightenlines,noshowpacs,nofootinbib,amsfonts,amsmath,amssymb,superscriptaddress,onecolumn]{revtex4-1}

\usepackage{amssymb,esvect,amsmath,graphicx,latexsym,amsthm,slashed,eso-pic,hyperref}

\usepackage{epsfig,amsmath,amssymb,verbatim,mathrsfs,hyperref}
\usepackage{xspace}
\usepackage{slashed}
\usepackage[utf8]{inputenc}
\graphicspath{{./}{Figs/}}

\def\beq{\begin{eqnarray}}
\def\eeq{\end{eqnarray}}
\def\bea{\begin{eqnarray}}
\def\eea{\end{eqnarray}}

\def\tev{\, {\rm TeV}}
\def\gev{\, {\rm GeV}}
\def\mev{\, {\rm MeV}}

\newcommand{\gsim}{\lower.7ex\hbox{$\;\stackrel{\textstyle>}{\sim}\;$}}
\newcommand{\lsim}{\lower.7ex\hbox{$\;\stackrel{\textstyle<}{\sim}\;$}}
\def\mpl{M_{\rm Pl}}

\newcommand{\ccdot}{\!\cdot\!}
\newcommand{\nnmb}{\nonumber}
\newcommand{\del}{\partial}

\newcommand{\lrf}[2]{\left(\frac{#1}{#2}\right)}

\newcommand{\zpp}{0^{++}}
\newcommand{\opm}{1^{+-}}

\begin{document}
\title{Cosmological Bounds on Non-Abelian Dark Forces}
\author{Lindsay Forestell}
\affiliation{TRIUMF, 4004 Wesbrook Mall, Vancouver, BC V6T 2A3, Canada}
\affiliation{Department of Physics and Astronomy, University of British Columbia, Vancouver, BC V6T 1Z1, Canada}
\author{David E. Morrissey}
\affiliation{TRIUMF, 4004 Wesbrook Mall, Vancouver, BC V6T 2A3, Canada}
\date{\today}
\author{Kris Sigurdson}
\affiliation{Department of Physics and Astronomy, University of British Columbia, Vancouver, BC V6T 1Z1, Canada}

\begin{abstract}
Non-Abelian \emph{dark} gauge forces that do not couple directly to 
ordinary matter may be realized in nature.  
The minimal form of such a dark force is a pure Yang-Mills theory.  
If the dark sector is reheated in the early universe, it will be realized 
as a set of dark gluons at high temperatures and as a collection of 
dark glueballs at lower temperatures, with a cosmological phase transition from 
one form to the other.  Despite being dark, the gauge fields of 
the new force can connect indirectly to the Standard Model through 
non-renormalizable operators.  These operators will transfer 
energy between the dark and visible sectors, and they allow some or 
all of the dark glueballs to decay.  In this work we investigate 
the cosmological evolution and decays of dark glueballs in the presence 
of connector operators to the Standard Model.  Dark glueball decays 
can modify cosmological and astrophysical observables, 
and we use these considerations to put very strong limits on the existence
of pure non-Abelian dark forces.  On the other hand, if one or more 
of the dark glueballs are stable, we find that they can potentially
make up the dark matter of the universe.

\end{abstract}

\maketitle


\section{Introduction\label{sec:intro}}

New gauge forces may be realized in nature beyond the 
$SU(3)_c\times SU(2)_L\times U(1)_Y$ structure of the Standard Model~(SM).  
If a new gauge force connects directly to SM matter, it must have a characteristic
mass scale above about a TeV to be consistent with experimental tests of 
the SM~\cite{Langacker:2008yv,Khachatryan:2016zqb,Aaboud:2017buh}.  
On the other hand, 
new \emph{dark} gauge forces that couple only very weakly to the SM can 
be significantly lighter~\cite{Pospelov:2008zw,Bjorken:2009mm,Essig:2013lka}.  
Such dark forces can be very challenging to probe directly in experiments,
and in many scenarios the strongest bounds on them come from astrophysical
and cosmological observations~\cite{Fradette:2014sza,Berger:2016vxi,Cohen:2016uyg,Fradette:2017sdd}.

In this work we investigate the cosmological evolution and constraints 
on new non-Abelian dark forces.  Such dark forces are well-motivated,
and arise in string theory constructions~\cite{Blumenhagen:2005mu}, 
models of dark matter or baryogenesis~\cite{Nussinov:1985xr,Barr:1990ca,Faraggi:2000pv,Gudnason:2006ug,Baumgart:2009tn,Kribs:2009fy,An:2009vq,Feng:2011ik,Cline:2013zca,Boddy:2014yra,Boddy:2014qxa,Yamanaka:2014pva,Garcia:2015loa,Soni:2016gzf,Kopp:2016yji,Ko:2016fcd,Dienes:2016vei,Halverson:2016nfq,Acharya:2017szw,Soni:2017nlm,Mitridate:2017oky}, neutral naturalness scenarios~\cite{Chacko:2005pe,Burdman:2006tz,Burdman:2014zta,Craig:2015pha,Curtin:2015fna},
and within the hidden valley paradigm~\cite{Strassler:2006im,Juknevich:2009ji,Juknevich:2009gg}.
The requirement of gauge invariance in theories of non-Abelian dark forces
implies that the new gauge vector bosons can only couple to the SM through
non-renormalizable operators~\cite{Juknevich:2009ji,Juknevich:2009gg}.  
This stands in contrast to Abelian dark forces that can connect to the SM 
at the renormalizable level through kinetic mixing with hypercharge.  
As a result, direct low-energy searches for non-Abelian dark forces are
very difficult, and cosmological observations usually provide the
most powerful tests of them~\cite{Feng:2011ik,Cline:2013zca,Boddy:2014yra,Boddy:2014qxa,Yamanaka:2014pva,Garcia:2015loa,Soni:2016gzf,Kopp:2016yji,Ko:2016fcd,Dienes:2016vei,Acharya:2017szw,Soni:2017nlm,Mitridate:2017oky,Halverson:2016nfq}.

The particle spectrum in theories of non-Abelian forces is diverse 
and complicated, and depends on both the gauge group 
and the representations of the matter fields charged under it.  
We focus on the minimal realization of a non-Abelian dark 
force consisting of a pure Yang-Mills theory with a simple gauge group $G_x$.  
Such theories can be described in terms of self-interacting dark gluons 
at high energies, but are expected to confine to a set of gauge-neutral 
glueball bound states below a dynamically generated confinement 
scale $\Lambda_x$~\cite{Jaffe:1985qp}.  
Both phases can be realized in the hot early universe, with a transition 
from the gluon phase to the glueball phase occurring
as the temperature (of the dark sector) falls below the critical 
temperature $T_c \sim \Lambda_x$~\cite{Lucini:2003zr}.  

  If the visible and dark sectors do not interact, they evolve
independently with distinct temperatures $T$ and $T_x$.  After confinement
at $T_x = T_c$, the dark glueballs undergo a complicated
freezeout process.  The energy density of the dark sector is dominated
by the lightest glueball state, which on general grounds is expected 
to have $J^{PC} = \zpp$~\cite{West:1995ym}.  The lightest $\zpp$ number density 
changes mainly through $(3\to 2)$ self-annihilation
processes~\cite{Soni:2016gzf,Forestell:2016qhc}.  
While these reactions are active, the dark temperature changes very slowly, 
only falling off as the logarithm of the cosmological scale 
factor~\cite{Carlson:1992fn,Dolgov:2017ujf}.  As a result,
the lightest glueballs form a massive thermal bath in which the other 
heavier glueballs annihilate through $2\to 2$ processes
and eventually freeze out~\cite{Pappadopulo:2016pkp,Forestell:2016qhc,Farina:2016llk}.  
In the end, a collection of relic glueball densities is left over, 
dominated by the $\zpp$ with exponentially smaller yields 
for the heavier states~\cite{Forestell:2016qhc,Farina:2016llk}.  

  The process of glueball freezeout can change drastically
if there are operators that connect the visible and dark sectors.
Such operators are always expected at some level; 
quantum gravitational effects are thought to induce gauge-invariant
operators involving both SM and dark sector fields suppressed by powers of
the Planck mass~\cite{Giddings:1988cx,Abbott:1989jw,Kallosh:1995hi,Banks:2010zn}.
Even stronger connections can arise if there exist new
matter fields that couple directly to both the visible and dark 
sectors~\cite{Juknevich:2009ji,Juknevich:2009gg}.
As long as the new physics generating these operators is much larger
than the confinement scale, their effects can be parametrized in terms
of a set of non-renormalizable connector operators.  

  With connectors, energy can be transferred between the dark and
visible sectors~\cite{Faraggi:2000pv,Soni:2016gzf,Forestell:2016qhc,Halverson:2016nfq,Acharya:2017szw}.  
After confinement, connector operators can also modify the glueball 
freeze-out dynamics and induce decays of some or all of the dark 
glueballs to the SM.  
If one of the glueballs is long-lived or stable, it will contribute
to the density of dark matter~(DM)~\cite{Boddy:2014yra}.  
However, glueball lifetimes that are not exceedingly long will inject
energy into the cosmological plasma and modify 
the standard predictions for big bang nucleosynthesis~(BBN)~\cite{
Kawasaki:2004qu,Jedamzik:2006xz} and 
the cosmic microwave background~(CMB)~\cite{Adams:1998nr,Chen:2003gz},
as well as act as astrophysical sources of cosmic and gamma rays~\cite{Cohen:2016uyg}.  

The aim of this work is to estimate the bounds on pure non-Abelian 
dark forces in the presence of connector operators from cosmology
and astrophysics.  
We focus mainly on the dark gauge group $G_x = SU(3)$ with glueball
masses above $m_0 \geq 100\,\mev$, and we study the leading connector 
operators between the dark vector bosons and the SM with characteristic 
mass scale $M\gg m_0$.  As an initial condition, we assume inflation 
(or something like it) followed by preferential reheating to the visible 
sector to a temperature above the confinement scale but below that of
the connectors.  With these assumptions, we find very strong limits 
on non-Abelian dark forces.

Cosmological effects of dark gluons and glueballs were studied previously
in Refs.~\cite{Faraggi:2000pv,Feng:2011ik,Boddy:2014yra,Boddy:2014qxa,Garcia:2015loa,Soni:2016gzf,Forestell:2016qhc,Halverson:2016nfq,Acharya:2017szw,Soni:2017nlm,Cohen:2016uyg}.  
We extend these earlier works with a more detailed
analysis of the leading (2-body) connector operators and their effects
on energy transfer between the visible and dark sectors.  We also investigate
the effects of heavier glueballs in the spectrum beyond the lightest mode,
and we show that the lightest $C$-odd glueball can play an important role
in some cases and even make up the observed DM density when it is long lived or stable.

Following this introduction, we discuss the general properties of glueballs 
relevant to our analysis in Sec.~\ref{sec:gb}.  Next, we present the
leading connector operators to the SM and investigate their implications
for glueball decays in Sec.~\ref{sec:decay}.  In Sec.~\ref{sec:gbfo} we
study the cosmological evolution of the dark gauge theory and we compute
glueball yields both with and without connector operators.  These results
are then applied to derive cosmological constraints on dark glueballs
in Sec.~\ref{sec:constraint}.  Finally, Sec.~\ref{sec:conclusions}
is reserved for our conclusions.
Some technical details about gluon thermalization and the cosmological
and astrophysical bounds we apply are collected 
in Appendices~\ref{sec:appa} and \ref{sec:appb}.

\section{Glueball Properties\label{sec:gb}}

    Glueballs have been studied using a variety of methods for a wide
range of non-Abelian gauge 
groups~\cite{Jaffe:1985qp,Mathieu:2008me}. 
In this section we review and derive some general results
for $SU(N)$ glueballs that will be essential for the analysis to follow.

\subsection{Glueball Masses}

  Detailed lattice studies of glueballs have been performed for $SU(N)$
with $N=2,\,3,\ldots$, and a number of stable states are found.
Since the minimal Yang-Mills action respects angular momentum~($J$), parity~($P$),
and charge conservation~($C$), glueballs can be classified according to 
their $J^{PC}$ quantum numbers.  The lightest state is found to have 
$J^{PC} = \zpp$~\cite{Morningstar:1999rf,Chen:2005mg}, 
as expected on general grounds~\cite{West:1995ym}.  The masses
and quantum numbers of the stable glueballs found for
$SU(2)$ and $SU(3)$ are listed in Tab.~\ref{tab:gb}.  They are
expressed in terms of the length scale $r_0$ where the
inter-gluon potential goes from Coulombic to linear, and for $SU(3)$
is related to the strong coupling scale by 
$r_0\Lambda_{x} = 0.614(2)(5)$~\cite{Gockeler:2005rv}.

\begin{table}[ttt]
\beq
\begin{array}{c|c|c}
J^{PC}&~m\,r_0\:(N=2)~&~m\,r_0\:(N=3)~\\
\hline\hline
0^{++}&4.5(3)&4.21(11)\\
2^{++}&6.7(4)&5.85(2)\\
3^{++}&10.7(8)&8.99(4)\\
0^{-+}&7.8(7)&6.33(7)\\
2^{-+}&9.0(7)&7.55(3)\\
\hline
1^{+-}&-&7.18(3)\\
3^{+-}&-&8.66(4)\\
2^{+-}&-&10.10(7)\\
0^{+-}&-&11.57(12)\\
1^{--}&-&9.50(4)\\
2^{--}&-&9.59(4)\\
3^{--}&-&10.06(21)\\
\end{array}
\nnmb
\eeq
\caption{Masses of known stable glueballs in $SU(2)$~\cite{Teper:1998kw}
and $SU(3)$~\cite{Morningstar:1999rf}.
\label{tab:gb}}
\end{table}

  For reasons to be explained below, we focus our attention on
two specific glueball states (for $SU(N\geq 3)$): the lightest overall
$J^{PC}=\zpp$ glueball in the spectrum, together with the lightest 
$C$-odd glueball with $J^{PC}=1^{+-}$.  With the gauge group $SU(3)$,
the mass of the lightest $\zpp$ glueball is 
$m_{0} \simeq 6.9\,\Lambda_x$, and the $\opm$ mass is 
$m_1 = 1.71(5)\,m_{0}$~\cite{Chen:2005mg}.

  Going beyond $SU(3)$ to larger $N$, the glueball mass spectrum
is found to be similar, with mass corrections suppressed by powers
of $1/N^2$~\cite{Teper:1998kw}.  Similar results are also expected for other
gauge groups with non-vanishing anomaly coefficient $d^{abc} = tr(t^a\{t^b,t^c\})$,
where $t^a$ is the generator of the fundamental 
representation~\cite{Jaffe:1985qp} (which we normalize according to
$tr(t^at^b) = \delta^{ab}/2$ for the $N$ of $SU(N)$).  
However, let us point out that 
for $SU(2)$ and other groups with vanishing $d^{abc}$ 
such as $SO(2N+1)$ and $Sp(2N)$, there are no $C$-odd states 
in the spectrum~\cite{Juknevich:2009ji}.  A further extension of 
the minimal Yang-Mills theory is the inclusion
of a topological theta term.  This would break $P$ and $T$, but not $C$.
It would also shift the glueball masses~\cite{Vicari:2008jw}, 
and induce mixing between glueball states with different 
$P$ quantum numbers~\cite{Vicari:2008jw,Gabadadze:2004jq}.

\subsection{Glueball Couplings and Matrix Elements \label{sec:selfinteractions}}

  Glueball self-couplings and transition matrix elements are needed
to compute their cosmological evolution.  These quantities have not been 
studied in as much detail on the lattice as the glueball mass spectrum.
Here, we collect the relevant existing lattice results, and we use 
naive dimensional analysis~(NDA)~\cite{Manohar:1983md,Cohen:1997rt,Nishio:2012sk} 
and large-$N$ scaling~\cite{tHooft:1973alw,Witten:1979kh} 
to make estimates when no lattice data is available.  

  Glueball interactions are expected to be perturbative in the
limit of large $N$ (for an underlying $SU(N)$ gauge group), and this
motivates writing an effective Lagrangian in terms of glueball fields. 
Combining the $N$ scaling of gluon $n$-point functions with dimensional
analysis suggests the form
\beq
\mathscr{L}_{eff} = \lrf{N}{4\pi}^2m_x^4\,F(\phi/m_x,\del/m_x) \ ,
\eeq  
where $\phi$ represents a glueball field interpolated by a single-trace
gluon operator, $m_x$ is a characteristic
glueball mass scale, and $F(x,y)$ is a smooth function that 
is finite as $N\to \infty$.
Expanding this function in a power series and rescaling
to obtain a canonical kinetic operator, the effective Lagrangian becomes
\beq
\mathscr{L}_{eff} = \frac{1}{2}(\del\phi)^2 
- \sum_n\frac{a_n}{n!}\,m_x^{4-n}\lrf{4\pi}{N}^{n-2}\phi^n + \ldots 
\label{eq:gbeft}
\eeq
where the coefficients $a_n$ are expected to be of order unity.
Note that shifting the gluon field to remove the linear term
does not alter this general form.
In the analysis to follow, we identify $m_x=m_0$ with the 
mass of the lightest glueball.

  We will also need glueball matrix elements in the analysis to follow.
Specific glueball states can be identified with gauge invariant gluon 
operators, in the sense that the operators can create one-particle
glueball states from the vacuum.  For example~\cite{Juknevich:2009ji},
\beq
\begin{array}{rclrl}
S&=& tr(X_{\mu\nu}X^{\mu\nu})&~\to~&\zpp\\
P&=& tr(X_{\mu\nu}\widetilde{X}^{\mu\nu})&~\to~&0^{-+}\\
T_{\mu\nu}&=& \frac{1}{2}tr(X_{\mu\alpha}X_{\nu}^{~\alpha}) - \frac{1}{4}\eta_{\mu\nu}S
&~\to~&2^{++},\,1^{-+},\,0^{++}\\
\Omega^{(1)}_{\mu\nu}&=&tr(X_{\mu\nu}X_{\alpha\beta}X^{\alpha\beta})
&~\to~&1^{--},\,1^{+-}
\label{eq:gops}\\
\Omega^{(2)}_{\mu\nu}&=&tr(X_{\mu}^{~\alpha}
X_{\alpha}^{~\beta}X_{\beta\nu})
&~\to~&1^{--},\,1^{+-}
\end{array}
\eeq
Here, $X_{\mu\nu} = X^a_{\mu\nu}t^a$ is the dark gluon field strength
contracted with the generators of the fundamental representation 
of the group normalized to $tr(t^at^b) = \delta^{ab}/2$.
  
  The two matrix elements of greatest interest to us are
\beq
\alpha_xF_{\zpp}^S &~\equiv~& 
\alpha_x\langle 0| tr(X_{\mu \nu}X^{\mu \nu}) | \zpp \rangle 
~\sim~ m_x^3
\label{eq:F0}
\\
\alpha_x^{3/2}M_{\opm \zpp} &~\equiv~& 
\alpha_x^{3/2}\langle \zpp | \big(\Omega_{\mu \nu}^{(1)}-\frac{5}{14}\Omega_{\mu \nu}^{(2)}\big)|\opm \rangle 
~\sim~ \sqrt{\frac{4\pi}{N}}\,m_x^3 \ ,
\label{eq:M10}
\eeq
where the estimates on the right hand sides are based on large-$N$ and NDA,
and $\alpha_x = g_x^2/4\pi$ is the dark gauge coupling.
In the second line, we have also suppressed the Lorentz structure of the
matrix element, $\epsilon_{\mu\nu\alpha\beta}\,p^{\alpha}\varepsilon^{\beta}$,
where $p^{\alpha}$ is the outgoing momentum and $\varepsilon^{\beta}$ is 
the polarization of the initial state~\cite{Juknevich:2009ji}.
The first of these matrix elements, $F_{\zpp}^S$, has been computed on the
lattice for $N=3$ with the result~\cite{Chen:2005mg,Meyer:2008tr}
\beq
4\pi\alpha_xF_{\zpp}^S = 2.3(5)\,m_x^3 \ ,
\eeq
which agrees reasonably well with our large-N and NDA estimate and is scale independent.  
In contrast, the second matrix element has not been calculated on the lattice.
We use the lattice value of $F_{\zpp}^S$ and the NDA estimate
$\alpha_x^{3/2}M_{\opm\zpp} = \sqrt{4\pi/N}\,m_x^3$ in the analysis to follow.

\section{Connections to the SM and Glueball Decays\label{sec:decay}}

  With the SM uncharged under the dark gauge group $G_x$, gauge invariance
forbids a direct renormalizable connection of the dark gluons to the SM. 
However, massive mediator states that couple to both sectors
can generate non-renormalizable operators connecting them.  If the characteristic
mass scale of the mediators is $M \gg \Lambda_x$, the leading 
operators have mass dimension of eight and six, 
and take the form~\cite{Juknevich:2009ji,Juknevich:2009gg}
\beq
\mathcal{O}^{(8a)} &~\sim~& \frac{1}{M^4}tr(F_{SM}F_{SM})\,tr(XX) \ ,
\label{eq:dim8a}\\
\mathcal{O}^{(8b)}&~\sim~&\frac{1}{M^4}B_{\mu\nu}\,tr(XXX)^{\mu\nu} \ ,
\label{eq:dim8b}\\
\mathcal{O}^{(6)}\phantom{i} &~\sim~& \frac{1}{M^2}H^{\dagger}H\,tr(XX) \ ,
\label{eq:dim6}
\eeq
where $X$ and $F_{SM}$ refer to the dark gluon and SM field strengths.
If present, these operators allow some or all of the glueballs
to decay to the SM.  In this section we illustrate simple mediator
scenarios that generate these operators, and we compute the glueball
decay rates they induce.

\subsection{Dimension-8 Operators}

  Dimension-8 operators of the form of Eqs.~(\ref{eq:dim8a},\ref{eq:dim8b})
lead to glueball decays with characteristic rate
\beq
\Gamma_8 ~\sim~ \frac{m_x^9}{M^8} \ .
\label{eq:d8approx}
\eeq
Here, we present an explicit scenario of mediator fermions that generates
these operators and we compute the glueball decay rates they induce.

  Before proceeding, it is helpful to organize the dimension-8 operators
according to a dark charge conjugation operation $C_x$ under which 
$X_{\mu}^a\to -\eta(a)X_{\mu}^a$, where $\eta(a)$ is the sign change of
the fundamental generator $t^a$ under charge conjugation~\cite{Peccei:1998jv}, 
with the SM vector bosons being invariant.  
The operators of Eq.~\eqref{eq:dim8a} are even under $C_x$ 
and those of Eq.~\eqref{eq:dim8b} are odd.  Furthermore, $C_x$
coincides with the $C_x$-number assignments of the glueball states.
Correspondingly, the operators of Eq.~\eqref{eq:dim8a} only allow 
direct decays of $C_x$-even glueballs to the SM, or glueball transitions
from even to even or odd to odd.  In particular, at $d=8$ the operator 
of Eq.~\eqref{eq:dim8b} is required for the lightest $C_x$-odd $1^{+-}$
glueball to decay.  

Consider now a set of massive vector-like fermions with masses 
$M_r\sim M \gg \Lambda_x$, each transforming as a fundamental 
or antifundamental under $G_x=SU(N)$ and the representation $r$  
of the SM gauge group (defined with respect to the left-handed component 
of the fermion).  Direct collider and precision electroweak limits on such 
fermions imply $M_r \gtrsim 100\,\gev$ if they only have electroweak charges,
and $M_r \gtrsim 1000\,\gev$ if they are charged under 
QCD~\cite{Juknevich:2009ji,Juknevich:2009gg,Lu:2017uur}.
The effective Lagrangian generated by integrating the fermions out 
is~\cite{Juknevich:2009ji}
\beq
\mathscr{L}_{eff} &\supset&
\frac{\alpha_x}{M^4}\left(
\alpha_1\chi_1B_{\mu\nu}B_{\alpha\beta}
+\alpha_2\chi_2W_{\mu\nu}^cW_{\alpha\beta}^c
+\alpha_3\chi_3G_{\mu\nu}^aG_{\alpha\beta}^a
\right)\nnmb\\
&&~\times\left(
\frac{1}{60}\,S\,\eta^{\mu\nu}\eta^{\alpha\beta}
+\frac{1}{45}\,P\,\epsilon^{\mu\nu\alpha\beta}
+\ldots
\right)\label{eq:leff8}\\
&&~+~\frac{\alpha_x^{3/2}\alpha_1^{1/2}}{M^4}\,\chi_Y\,B_{\mu\nu}\,
\frac{14}{45}\left(\Omega^{(1)}_{\mu\nu}-\frac{5}{14}\Omega^{(2)}_{\mu\nu}\right)
\ .
\eeq
Here, the dark gluon operators $S,\,P,$ and $\Omega^{(1,2)}_{\mu\nu}$ correspond
to Eq.~\eqref{eq:gops}, and the coefficients $\chi_i$ 
are given by
\beq
\chi_i &=& \sum_rd(r_i)T_{2}(r_i)/\rho_r^4
\label{eq:chii}\\
\chi_Y &=& \sum_rd(r_i)Y_r/\rho_r^4\ ,
\label{eq:chiy}
\eeq
where the sums run over the SM representations $r$ of the fermions,
and $\rho_r = M_r/M$.
For each such representation, we define sub-representations 
$r=(r_1,r_2,r_3)$ with respect to the SM gauge factors 
$G_i = U(1)_Y,\,SU(2)_L,\,SU(3)_c$.  The quantity $d(r_i)$ is the number 
of copies of the $i$-th sub-representation within $r$,
and $T_2(r_i)$ is the trace invariant for that factor 
(normalized to $1/2$ for the $N$ of $SU(N)$ and $Y^2$ for $U(1)_Y$).\footnote{
Note that our $\chi_{2,3}$ are smaller by a factor of $1/2$ than the
corresponding terms in Ref.~\cite{Juknevich:2009ji}.}

  Generic representations of mediator fermions break the dark
charge conjugation number $C_x$ explicitly and generate both operator types 
of Eqs.~(\ref{eq:dim8a},\ref{eq:dim8b}).
This is explicit in Eq.~\eqref{eq:leff8}, with both even~($\chi_i\neq 0$) 
and odd operators~($\chi_Y\neq 0$).  However, there exist mediator fermion
combinations that preserve $C_x$~\cite{DiFranzo:2015nli} and yield $\chi_Y = 0$.
From Eq.~\eqref{eq:chiy}, we see that this requires a specific combination
of fermion charges as well as masses.  The presence of masses also implies
that $C_x$ can be broken softly.  In contrast, the $\chi_i$ coefficients
of Eq.~\eqref{eq:chii} are positive semi-definite and not subject to 
cancellation.  

  The $C_x$-preserving operator of Eq.~\eqref{eq:leff8} allows direct
decays of the $\zpp$ glueball to pairs of SM vector bosons.  The corresponding
decay widths are~\cite{Juknevich:2009ji}
\beq
\Gamma(\zpp\to gg) &~=~& (N_c^2-1)\frac{\alpha_3^2}{16\pi}\lrf{2}{60}^2
\!\chi_3^2\,
\frac{m_0^3(\alpha_xF_{\zpp}^S)^2}{M^8} \ ,
\label{eq:sglue}
\\
\frac{\Gamma(\zpp\to \gamma\gamma)}{\Gamma(\zpp\to gg)} &~=~& 
\frac{1}{(N_c^2-1)}\lrf{\alpha\chi_{\gamma}}{\alpha_3\chi_3}^2\\
\frac{\Gamma(\zpp\to ZZ)}{\Gamma(\zpp\to gg)} &~=~& 
\frac{1}{(N_c^2-1)}\lrf{\alpha_2\chi_Z}{\alpha_3\chi_3}^2
\left(1-4\frac{m_Z^2}{m_0^2}\right)^{1/2}
\left(1-4\frac{m_Z^2}{m_0^2}+6\frac{m_Z^4}{m_0^4}\right)
\\
\frac{\Gamma(\zpp\to W^+W^-)}{\Gamma(\zpp\to gg)} &~=~& 
\frac{2}{(N_c^2-1)}\lrf{\alpha_2\chi_2}{\alpha_3\chi_3}^2
\left(1-4\frac{m_W^2}{m_0^2}\right)^{1/2}
\left(1-4\frac{m_W^2}{m_0^2}+6\frac{m_W^4}{m_0^4}\right)
\\
\frac{\Gamma(\zpp\to \gamma Z)}{\Gamma(\zpp\to gg)} &~=~& 
\frac{2}{(N_c^2-1)}\lrf{\sqrt{\alpha\alpha_2}\,\chi_{\gamma Z}}{\alpha_3\chi_3}^2
\left(1-\frac{m_Z^2}{m_0^2}\right)^3 
\eeq
where $m_0 = m_x$ is the $\zpp$ glueball mass, 
$F_{\zpp}^S$ is given by Eq.~\eqref{eq:F0}, 
$N_c^2-1 =8$,
the $\chi_i$ are defined in Eq.~\eqref{eq:chii}, 
$\chi_\gamma = \chi_1 + \chi_2$, 
$\chi_Z = (s_W^4\chi_1+c_W^4\chi_2)/c_W^2$, 
and $\chi_{\gamma Z} = (c_W^2\chi_2 - s_W^2\chi_1)/c_W$, 
with $s_W$ being the sine of the weak mixing angle.  
Note that the decay width to gluons in Eq.~\eqref{eq:sglue} only applies
for $m_0 \gg 1\,\gev$; at lower masses the final states consist of
hadrons.  We do not attempt to model this hadronization, and instead
we apply a factor of $\sqrt{1-(2m_{\pi}/m_0)^2}$ to the decay width. 
In evaluating the width of Eq.~\eqref{eq:sglue}, we take $\alpha_3$
at scale $m_0$ since the corresponding gluon operator is renormalized
(at one-loop) in the same way as the standard field strength operator.

  Decays of the lightest $1^{+-}$ glueball occur through the 
$C_x$-odd operator term in Eq.~\eqref{eq:leff8}, with the leading decay 
channels expected to be $\opm\to \zpp+\{\gamma,Z\}$.  
The widths are~\cite{Juknevich:2009ji}
\beq
\Gamma(1^{+-}\to\zpp\gamma) &=& 
\frac{\alpha}{24\pi}\,\chi_Y^2\left(1-\frac{m_{x}^2}{m_{1}^2}\right)^3
\frac{m_{1}^3\,(\alpha_x^{3/2}M_{1^{+-}\zpp})^2}{M^8} \\
\Gamma(1^{+-}\to\zpp Z) &=& 
\frac{\alpha}{24\pi}t_W^2\chi_Y^2
\left[\left(1+\frac{m_{x}^2}{m_1^2}-\frac{m_Z^2}{m_1^2}\right)^2
-4\,\frac{m_{x}^2}{m_{1}^2}\right]^{3/2}
\frac{m_{1}^3\,(\alpha_x^{3/2}M_{1^{+-}\zpp})^2}{M^8} 
\eeq
with $m_1 = m_{\opm}$, and $M_{\opm\zpp}$ defined in Eq.~\eqref{eq:M10}.

\begin{figure}[ttt]
  \centering
	\includegraphics[width=0.35\textwidth]{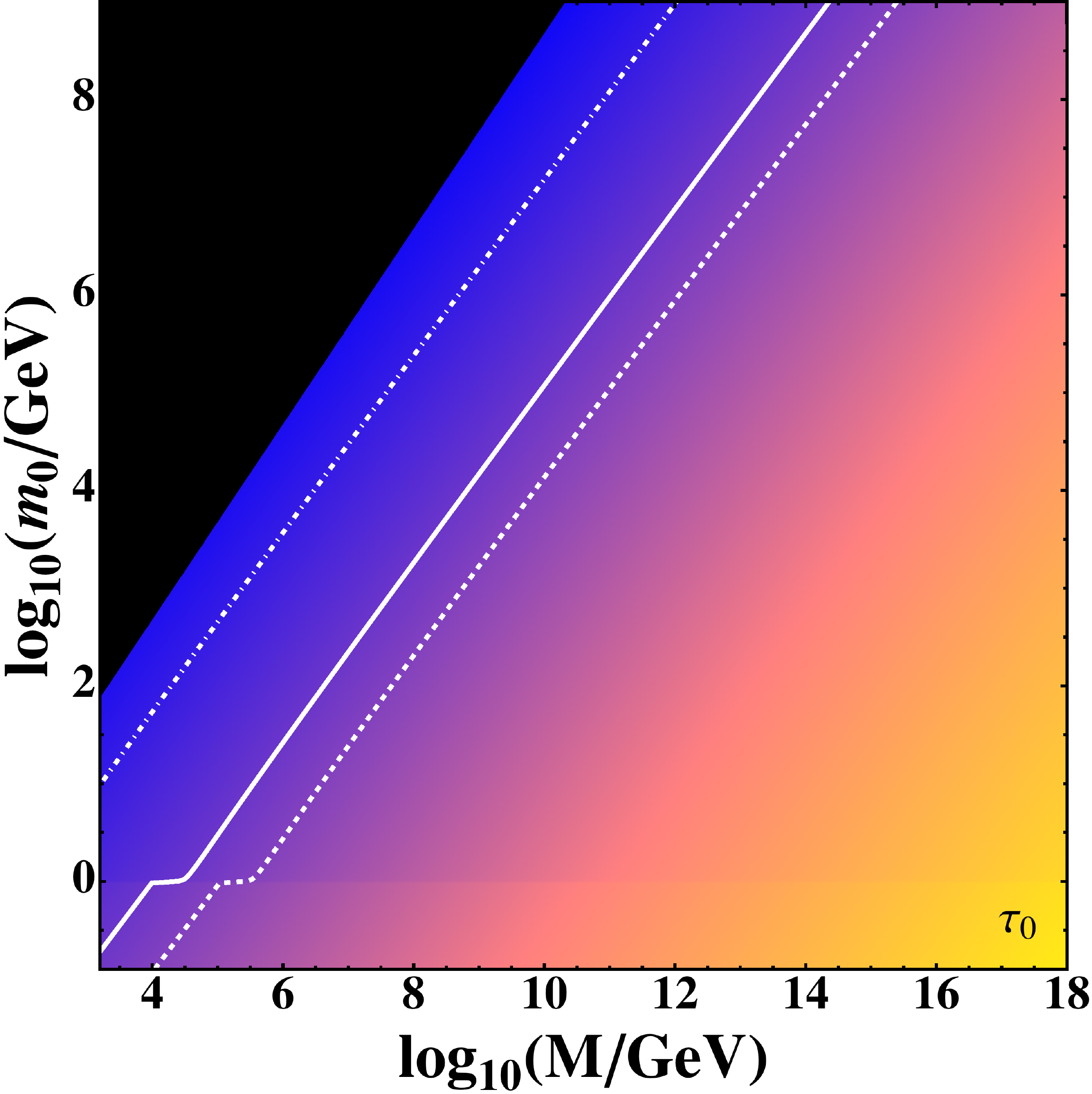}
	\includegraphics[width=0.35\textwidth]{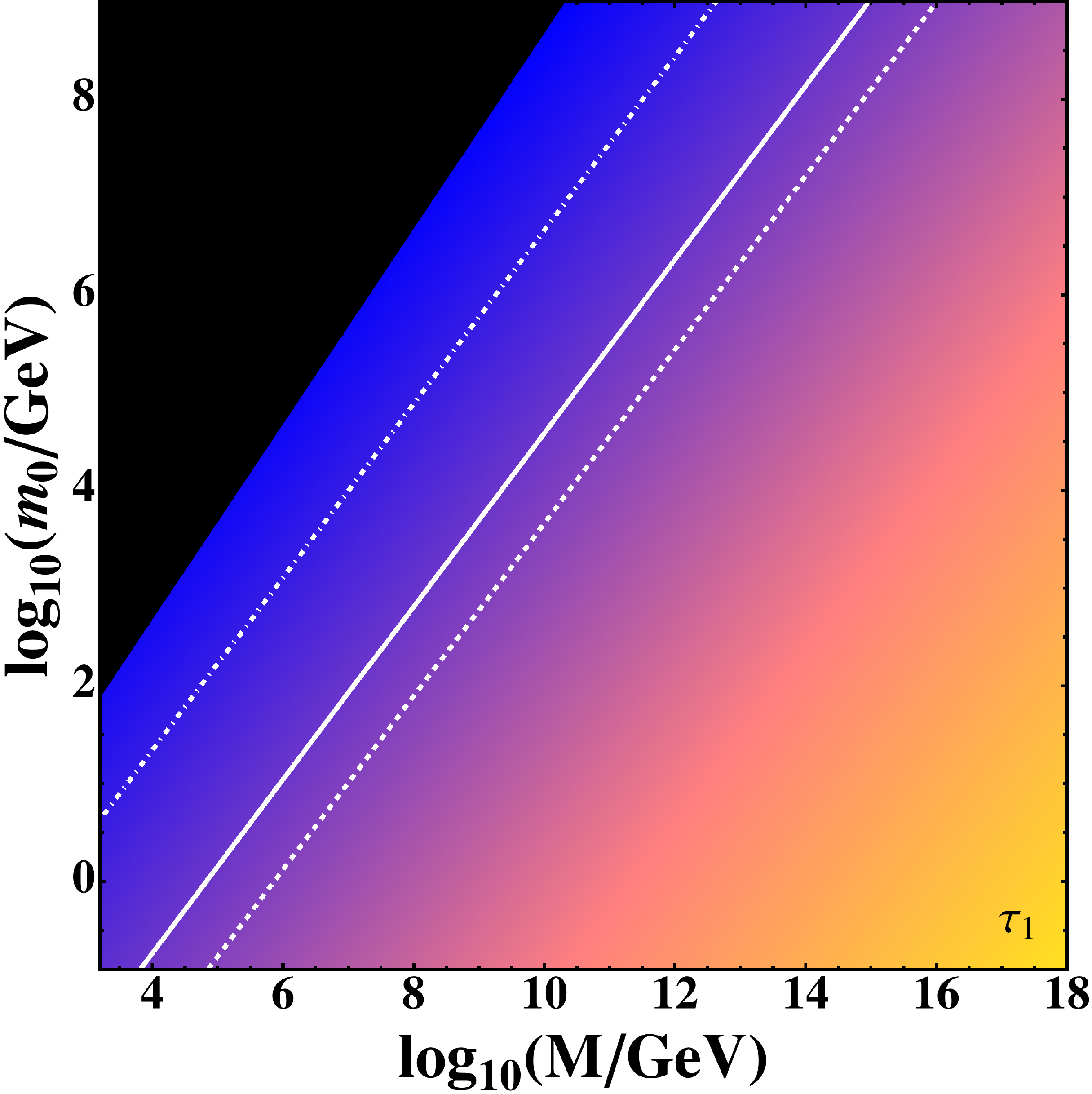}
	\raisebox{1.0cm}{\includegraphics[width=0.2\textwidth]{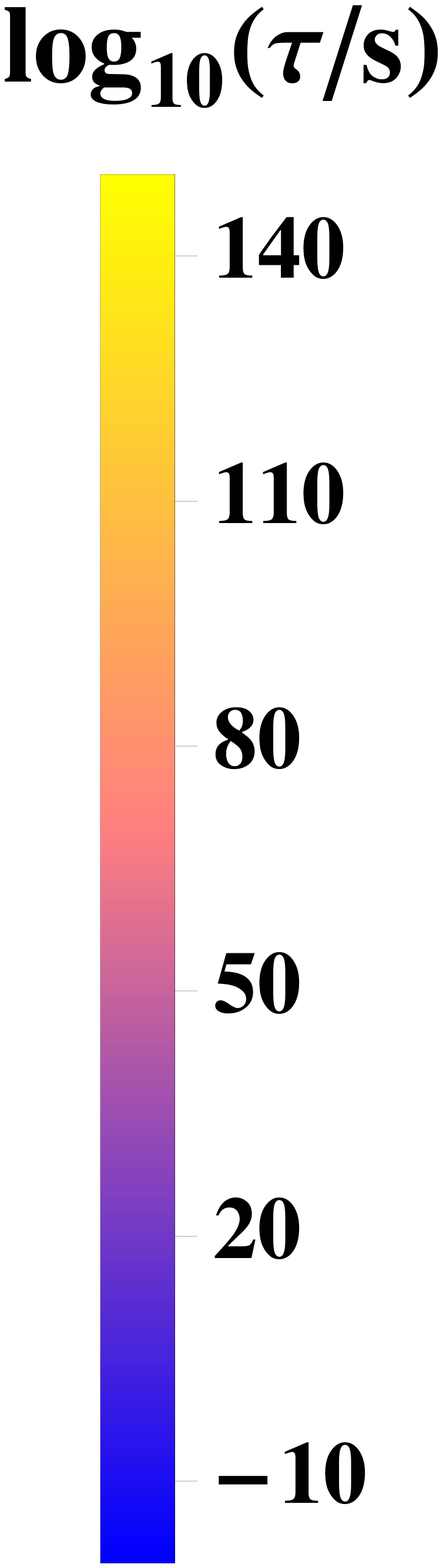}}
\caption{Decay lifetimes $\tau=1/\Gamma$ of the $\zpp$~(left) and $\opm$~(right)
glueball states due to the dimension-8 operators as a function of $M$ and $m_0$ 
for $\chi_i=\chi_Y=1$ and $G_x=SU(3)$. The masked regions at the upper left 
show where $m_0 > M/10$ and our treatment in terms of effective operators 
breaks down, while the white dotted, solid, and dashed lines indicate 
reference lifetimes of 
$\tau = 0.1\,\text{s},~5\times 10^{17}\,\text{s},~10^{26}\,\text{s}$. 
\label{fig:decrates8}}
\end{figure}

  The total decay lifetimes $\tau=1/\Gamma$ of the $\zpp$ and $\opm$ 
glueball states from the dimension-8 operators above with 
$\chi_i=\chi_Y=1$ and $G_x=SU(3)$ are shown in the left and right panels 
of Fig.~\ref{fig:decrates8}.  In the upper left of both plots, we mask out 
the regions with $m_0 > M/10$ where our treatment in terms of 
effective operators breaks down.  
The dotted, solid, and dashed lines indicate reference lifetimes 
of $\tau = 1/\Gamma = 0.1\,\text{s},~5\times 10^{17}\,\text{s},10^{26}\,\text{s}$. These lifetimes correspond to decays that occur early in the history of the Universe, at the present day, and long lived glueballs, respectively.
Both decay rates follow the approximate scaling of Eq.~\eqref{eq:d8approx}.  
All other known ($SU(3)$) glueballs can decay through these dimension-8
operators as well with parametrically similar rates, although there can be
numerically significant differences due to coupling factors 
and phase space~\cite{Juknevich:2009ji}.

\subsection{Dimension-6 Operators}

  Glueball decays through the dimension-6 operator of Eq.~\eqref{eq:dim6}
proceed with characteristic rate
\beq
\Gamma_6 ~\sim~ \frac{m_0^5}{M^4} \ .
\label{eq:dec6}
\eeq
We present here two mediator scenarios that generate the 
operator of Eq.~\eqref{eq:dim6} and we compute the decay rates
they induce.

  Our first mediator scenario follows Ref.~\cite{Juknevich:2009gg} 
and consists of mediator fermions with Yukawa couplings to the SM Higgs boson. 
A minimal realization contains a vector-like $SU(2)_L$ doublet $P$ 
with gauge quantum numbers $(r_x,1,2,-1/2)$, and a vector-like singlet 
$N$ with quantum numbers ($r_x,1,1,0$) together with 
the interactions~\cite{Juknevich:2009gg,Lu:2017uur}
\beq
-\mathscr{L} ~\supset~ M_P\bar{P}P + M_N \bar{N}N + \lambda\bar{P}HN + (h.c.) \ .
\eeq
For $M_N,\,M_P\gg m_h$, the leading glueball effective operator from
integrating out the fermions can be obtained using the low-energy Higgs
theorem~\cite{Shifman:1979eb},
\beq
\mathscr{L}_{eff} ~\supset~ \frac{\alpha_x}{6\pi} T_2(r) \frac{\lambda^2}{M^2} H^\dagger H X^{a}_{\mu \nu}X^{a~\mu \nu} \ ,
\eeq
where $M^2\simeq M_PM_N$ and $T_2(r_x)=1/2$ is the trace invariant of the fermion 
representation $r_x$ under the dark gauge group $G_x$.  
In addition to the dimension-6 operator above, the massive fermions 
also generate dimension-8 operators of the form of Eq.~\eqref{eq:leff8}.  

  A second mediator scenario consists of a complex scalar $\Phi_x$ charged
under the dark gauge group with a Higgs-portal coupling,
\beq
-\mathcal{L} \supset M_\Phi^2 |\Phi_x|^2+\kappa |\Phi_x|^2|H|^2
\label{eq:hport}
\eeq
Applying the low-energy Higgs theorem to this state (for $M_\Phi \gg m_h$),
we find
\beq
-\mathscr{L}_{eff} ~\supset~ 
-\frac{\alpha_x}{48\pi}T_2(r)\frac{\kappa}{M_{\Phi}^2}\,
H^{\dagger}H\,X_{\mu\nu}^aX^{a\,\mu\nu} \ .
\eeq
In passing, we note that the Higgs portal coupling of Eq.~\eqref{eq:hport}
respects dark $C_x$ number.  

  The operator generated in either mediator scenario can be written
in the form
\beq
-\mathscr{L}_{eff} ~\supset~  
\frac{\alpha_x y_{eff}^2}{6\pi M^2} H^\dagger H X^a_{\mu \nu} X^{a\,\mu \nu} \ ,
\label{eq:leff6}
\eeq
with the dimensionless coefficient $y_{eff}$.  
Since this operator is even under $C_x$, it only allows direct decays of 
$C_x$-even glueballs to the SM, or even-to-even or odd-to-odd glueball
transitions.  It was shown in Ref.~\cite{Juknevich:2009gg} that this is sufficient
to allow all known $SU(3)$ glueballs to decay, except for the $\opm$
and $0^{-+}$ modes.  The absence of a $\opm$ decay follows from $C_x$
considerations, while the conclusion for $0^{-+}$ is a result of 
spin and parity, rather than $C_x$.  This mode can decay at the dimension-6
level if a topological dark gluon term is added to the UV Lagrangian
or by extending to a two-Higgs doublet model~\cite{Juknevich:2009gg}.

  Using the parametrization of Eq.~\eqref{eq:leff6}, the direct decay of 
the $\zpp$ glueball to the SM has rate~\cite{Juknevich:2009gg}
\beq
\Gamma(\zpp\to SM) = 
\lrf{y_{eff}^2}{3\pi}^2
\frac{(\sqrt{2}\langle H\rangle)^2\,(\alpha_xF_{\zpp}^S)^2}{M^4\,[(m_0^2-m_h^2)^2+(m_h\Gamma_h)^2]}\,
\Gamma_h(m_h\to m_0) \ ,
\label{eq:gam6}
\eeq
where $\sqrt{2}\langle H\rangle = 246\,\gev$ is the electroweak vacuum 
expectation value, 
$F_{\zpp}^S$ is defined in Eq.~\eqref{eq:F0}, $m_h = 125\,\gev$ 
is the Higgs mass, $\Gamma_h=4.1\,\mev$ is the Higgs width, 
and $\Gamma_h(m_h\to m_0)$ is the total width the SM Higgs would have 
if its mass were $m_0$ (and includes decays to Higgs final states 
for $m_0 > 2m_h$).  We evaluate this width using the expressions 
of Refs.~\cite{Gunion:1989we,Gunion:1992hs}.

\begin{figure}[ttt]
	\centering
	\includegraphics[width=0.45\textwidth]{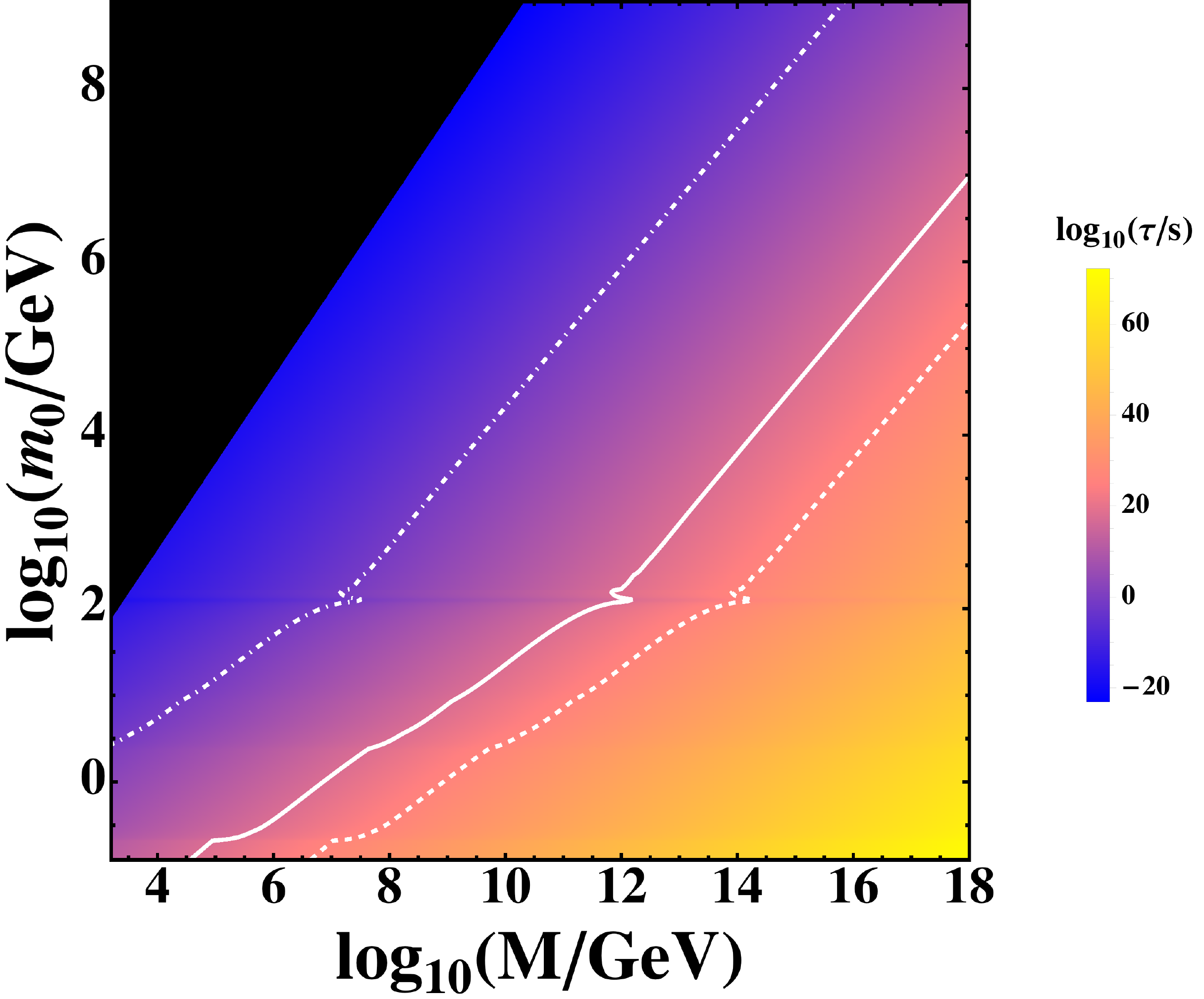}
	\caption{Decay lifetime $\tau=1/\Gamma$ of the $\zpp$ 
glueball due to the combined dimension-6 and dimension-8 operators 
as a function of $M$ and $m_0$ for $\chi_i=\chi_Y=1$, $y_{eff}=1$, 
and $G_x=SU(3)$.  The masked region at the upper left shows where 
$m_0 > M/10$ and our treatment in terms of effective operators breaks down, 
while the dotted, solid, and dashed white lines indicate lifetimes of 
$\tau$ = $0.1\,\text{s},~5\times 10^{17}\,\text{s},~10^{26}\,\text{s}$.
\label{fig:decrates6}}
\end{figure}

  In Fig.~\ref{fig:decrates6} we show the decay lifetime
$\tau=1/\Gamma$ of the $\zpp$ glueball from the dimension-6 (and dimension-8)
operators above with $y_{eff}=1$ and $G_x=SU(3)$. 
The upper region of the plot is masked out since it corresponds to $m_0 > M/10$ 
where our treatment in terms of effective operators breaks down.  
The dotted, solid, and dashed lines indicate lifetimes 
of $\tau = 0.1\,\text{s},~5\times 10^{17}\,\text{s},10^{26}\,\text{s}$.
For $m_0 \gg m_h$, the $\zpp$ lifetime scales according to Eq.~\eqref{eq:dec6},
while for $m_0 < m_h$ there is an additional suppression from small Yukawa
couplings.  Comparing to the $\opm$ lifetime in Fig.~\ref{fig:decrates8},
we see that it is parametrically long-lived compared to the $\zpp$ when
both dimension-6 and dimension-8 operators are present.

\subsection{Decay Scenarios}
\label{sec:dscenario}

  Based on the discussion above, we present four glueball decay
scenarios organized by the dimensions of the relevant decay operators
and the dark conjugation charge $C_x$: 
\begin{enumerate}
\item \textbf{Dimension-$8$ decays with broken $C_x$}\vspace{0.1cm}\\
In this scenario glueballs decay exclusively through the dimension-8
operators of the form of Eq.~\eqref{eq:leff8}.  
All glueballs are able to decay with parametrically similar rates.
To realize this scenario, we use the effective interactions 
in Eq.~\eqref{eq:leff8} with $\chi_i=\chi_Y=1$.  
\item \textbf{Dimension-$8$ decays with exact $C_x$}\vspace{0.1cm}\\
This scenario is similar to the first, but now with $\chi_Y=0$.
Conservation of $C_x$ implies that the lightest $\opm$ glueball is stable.
The other glueballs are all able to decay with parametrically similar rates.
\item \textbf{Dimension-$6$ decays with broken $C_x$}\vspace{0.1cm}\\
Glueball decays occur through the dimension-6 operator 
of Eq.~\eqref{eq:leff6} and the dimension-8 operators 
of Eq.~\eqref{eq:leff8}.  We realize the scenario by setting 
$y_{eff}=1$ together with $\chi_i=\chi_Y=1$.
With the exception of the $\opm$ mode (and possibly the $0^{-+}$),
glueballs decay primarily through the dimension-6 operator.  
In contrast, the $\opm$ glueball only decays
through the $C_x$-breaking dimension-8 operator with a parametrically
suppressed rate, making it much longer-lived than the other glueballs, which in turn leads to different cosmological scenarios when considering the constraints we can place on this model.

\item \textbf{Dimension-$6$ decays with exact $C_x$}\vspace{0.1cm}\\
Decays occur through the dimension-6 operator of Eq.~\eqref{eq:leff6} 
and the $C_x$-conserving terms in Eq.~\eqref{eq:leff8}.
We realize the scenario by taking $y_{eff}=1$, $\chi_i=1$, and $\chi_Y=0$.
The $\opm$ glueball is stable, while the other glueballs decay mainly
through the dimension-6 operator.
\end{enumerate}
We study the cosmological implications of
these four decay scenarios in the analysis to follow.

\section{Glueball Densities in the Early Universe\label{sec:gbfo}}

  Glueballs are formed in the early universe in a confining transition
as the dark sector temperature $T_x$ falls below a critical temperature
$T_c \sim m_0$.  After they are created, the glueballs undergo 
a complicated freezeout process involving a range of $2\to 2$ and $3\to 2$ 
reactions.  These dynamics become even more complicated when the 
dark sector connects to the SM through the operators discussed above,
with new effects such as energy transfer between the visible and dark sectors
and glueball decays. In this section we review the formation and freezeout 
of glueballs in the absence of connectors to the SM, and we investigate how 
this picture changes when connectors are present.

\subsection{Glueball Formation and Freezeout without Connectors}
\label{sec:gbform}

  In the absence of operators that connect to the SM, the visible and dark sectors
do not thermalize with each other.  We assume that enough energy is
liberated by reheating following primordial inflation (or something similar)
that both sectors are able to thermalize independently with temperatures 
$T$ and $T_x$~\cite{Adshead:2016xxj},
and furthermore that $T_x \geq T_c$ at this point.\footnote{
If not, the glueball relic density is set by the details of inflationary
reheating.}

As the universe expands and cools, dark glueballs are formed in 
a confining transition.  This transition has been studied in detail 
using lattice methods for $G_x = SU(N)$~\cite{Boyd:1996bx,Lucini:2003zr,Lucini:2005vg,Lucini:2012wq,Borsanyi:2012ve}, 
and the critical temperature of the transition is found to be~\cite{Lucini:2012wq}
\beq
T_c\,r_0 = 0.709(6) + 0.546(22)/N^2 \ ,
\eeq
corresponding to $T_c \simeq m_0/5.5$ for $N=3$.\footnote{We have 
used the results of Ref.~\cite{Guagnelli:1998ud} to
convert between different lattice conventions.}
The confining transition is also found to be second-order for $N=2$,
very weakly first-order for $N=3$, and increasingly strongly
first-order for larger $N$~\cite{Lucini:2003zr,Lucini:2005vg}.
Generalizing the analysis of Ref.~\cite{Garcia:2015loa} as in
Ref.~\cite{Forestell:2016qhc}, we estimate that the fractional
entropy change in the transition is negligible for $N\lesssim 10$
over the full range of parameters considered in this work.
Following the transition, as $T_x$ falls below the critical temperature, 
the glueball masses quickly settle 
to their zero-temperature values~\cite{Ishii:2002ww,Meng:2009hh}.
Based on these results, we assume the phase transition occurs 
instantaneously at $T_x=T_c$.  This should be a good approximation
for $N=3$, but could be inaccurate for much larger $N$.  

Entropy is conserved independently in both sectors while kinetic equilibrium
is maintained.  This implies that the ratio of entropy densities $s$ and $s_x$ 
in the two sectors remains constant,
\beq
R ~\equiv~ \frac{s_x}{s}~=~ constant \ . 
\eeq
We take $R$ as an input to our calculation; in the absence of connectors
its value is set by the unspecified dynamics of reheating after 
inflation~\cite{Adshead:2016xxj}.
However, we do assume $R < 1$ corresponding to preferential reheating
to the visible sector.  For $T_x \gg T_c$ and $G_x=SU(N)$, 
the entropy ratio is related to the temperatures in the two sectors by
\beq
R = \frac{2(N^2-1)}{g_{*S}}\lrf{T_x}{T}^3 \ ,
\eeq
where $g_{*S}$ is an effective number of degrees of freedom
in the visible sector at temperature $T$.  This ratio will be
maintained through the confining transition provided it is not
too strongly first order~\cite{Garcia:2015loa}.

  Once formed, dark glueballs interact with each other
and undergo a freezeout process in which they depart from thermodynamic
equilibrium and develop stable relic densities.  This process was studied 
in detail in Refs.~\cite{Garcia:2015loa,Soni:2016gzf,Forestell:2016qhc}.  
In the last work, the evolution of glueball numbers was computed numerically 
using a network of Boltzmann equations containing the most important 
$2\to 2$ and $3\to 2$ reactions, with thermally averaged cross sections 
estimated using the glueball effective Lagrangian of Eq.~\eqref{eq:gbeft}.  
  
  For the purposes of our cosmological analysis of glueball effects to follow,
the results of Ref.~\cite{Forestell:2016qhc} are captured to an excellent 
approximation by a simplified two-state model for the densities of the $\zpp$ 
and $\opm$ glueballs.  In this model, the evolution equations
for the $\zpp$ density $n_0$ and the $\opm$ density 
$n_1$ are~\cite{Forestell:2016qhc}
\beq
\frac{dn_0}{dt} +3Hn_0 &~=~&
-\langle\sigma_{32}v^2\rangle n_0^2(n_0-\bar{n}_0)
\label{eq:evol0}\\
&&~~+~\langle\sigma_{22} v\rangle\left[n_1^2-\lrf{n_0}{\bar{n}_0}^2\!\bar{n}_1^2
\right] \ ,
\nnmb\\
\frac{dn_1}{dt} +3Hn_1 &~=~&
-\langle\sigma_{22} v\rangle\left[n_1^2-\lrf{n_0}{\bar{n}_0}^2\!\bar{n}_1^2\right] \ ,
\label{eq:evol1}
\eeq
where $H$ is the Hubble factor, $\bar{n}_i$ are the equilibrium number 
densities at temperature $T_x$, and $\langle\sigma_{32}v\rangle$ 
and $\langle\sigma_{22} v\rangle$ correspond to the $3(\zpp)\to 2(\zpp)$ 
and $(\opm\opm) \to (\zpp\zpp)$ processes.
In detail, the Hubble factor is given by
\beq
H^2 = \frac{1}{3\mpl^2}(\rho+\rho_x) \ ,
\label{eq:hubble}
\eeq
where $\rho$ is the energy density of the SM and $\rho_x$ is that of
the dark sector.  Since kinetic equilibrium is expected to hold 
in the dark sector throughout the freezeout process,
the number densities take the form
\beq
n_i = g_i\int\!\frac{d^3p}{(2\pi)^3}\left[e^{(E_i-\mu_i)/T_x}-1\right]^{-1} \ ,
\eeq
where $E_i=\sqrt{m_i^2+\vec{p}^2}$, $g_i$ is the number of internal
degrees of freedom, $\mu_i$ is a chemical potential,
and $T_x$ is the common dark sector temperature.  The precise definition 
of the equilibrium densities is then $\bar{n}_i = n_i(T_x,\,\mu_i\to 0)$.  
For the thermally averaged cross sections, we estimate them
using Eq.~\eqref{eq:gbeft}:
\beq
\langle\sigma_{32}v^2\rangle &~\simeq~& 
\frac{1}{(4\pi)^3} \lrf{4\pi}{N}^6\frac{1}{m_0^5} \ ,
\\
\langle\sigma_{22} v\rangle &~\simeq~& \frac{1}{4\pi}\lrf{4\pi}{N}^4
\frac{1}{m_1^2} \ .
\eeq

To evaluate Eqs.~(\ref{eq:evol0},\ref{eq:evol1}), it is necessary to track 
the time evolution of the dark temperature $T_x$.
This can can achieved using the constancy of the entropy ratio $R$ together with 
\beq
T_xs_x = \sum_i(\rho_i+p_i-\mu_in_i)  \ .
\label{eq:entro} 
\eeq
Prior to freezeout of the $\zpp$ mode (and after dark confinement), 
the dark temperature falls as $T_x \propto 1/\ln(a)$ due to the energy 
injected by $3\to 2$ annihilations~\cite{Carlson:1992fn}.  
After $\zpp$ freezeout, the dark temperature falls as $T_x \propto a^{-2}$.  

  Our two-state simplified model provides an excellent approximation 
of the full analysis of Ref.~\cite{Forestell:2016qhc}.  
A central feature of the analysis is that near equilibrium at $T_x < T_c$ 
the $2\to 2$ reactions are parametrically faster than the $3\to 2$.
This keeps the ratio of the $\zpp$ and $\opm$ densities close to the equilibrium
ratio throughout the freezeout process,
\beq
\frac{n_1}{n_0} ~\simeq~ \frac{\bar{n}_1}{\bar{n}_0} 
~=~ 3\lrf{m_1}{m_0}^{3/2}e^{-x_x(\Delta m/m_0)} \ ,
\eeq
with $\Delta m = (m_1-m_0)$ and $x_x = m_0/T_x$. 
Thus, the $2\to 2$ reactions push the $\opm$ density to be
exponentially smaller than the $\zpp$.  

The exponential suppression of heavier modes also means that  
the freezeout of the $\zpp$ mode can be computed reliably in isolation,
neglecting the effects of the $\opm$ and keeping only the $3\to 2$ reactions.  
Calculations of single-state freezeout through $3\to 2$ annihilation
have been performed in Refs.~\cite{Carlson:1992fn,Pappadopulo:2016pkp,Soni:2016gzf,Farina:2016llk}.   For freezeout at $x_x^{fo} = T_x^{fo}/m_0$, 
the $\zpp$ relic yield is approximately
\beq
Y_0 ~\equiv~ \frac{n_0}{s} ~\simeq~ \frac{R}{x_x^{fo}} \ .
\eeq
Numerically, we find $x_x^{fo} \in [5,20]$ for $R\in [10^{-12},0.1]$
and $m_0\in [10^{-3},10^{9}]\,\gev$.  
The less-abundant $\opm$ mode freezes out in the background of the
massive bath of $\zpp$ glueballs~\cite{Forestell:2016qhc,Farina:2016llk}. 
This occurs after $\zpp$ freezeout, but before the kinetic 
self-equilibration of the $\zpp$ states is lost.  

\begin{figure}[ttt]
  \centering
  \includegraphics[width=0.35\textwidth]{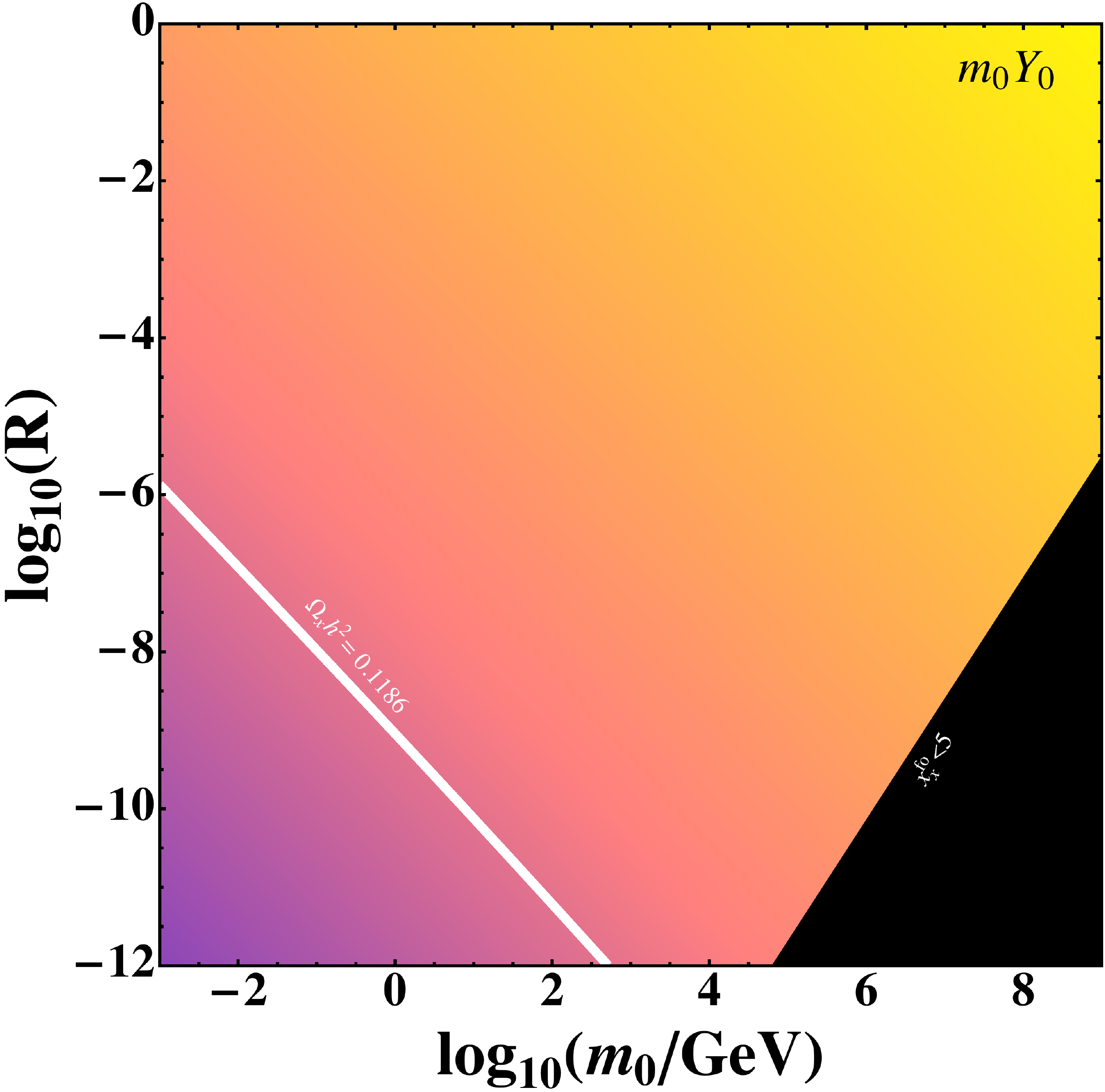}
  \includegraphics[width=0.35\textwidth]{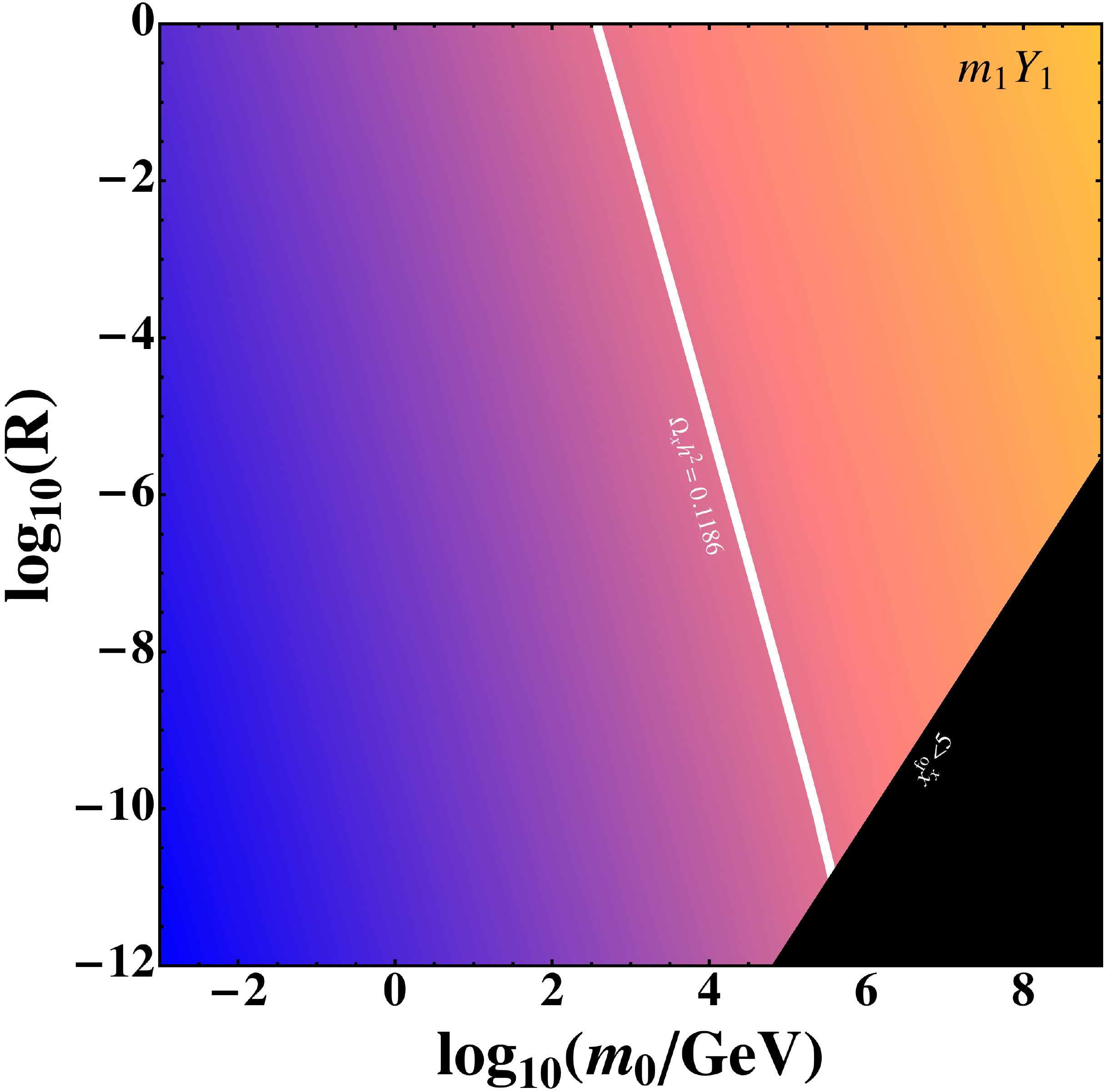}
  \raisebox{1cm}{\includegraphics[width=0.2\textwidth]{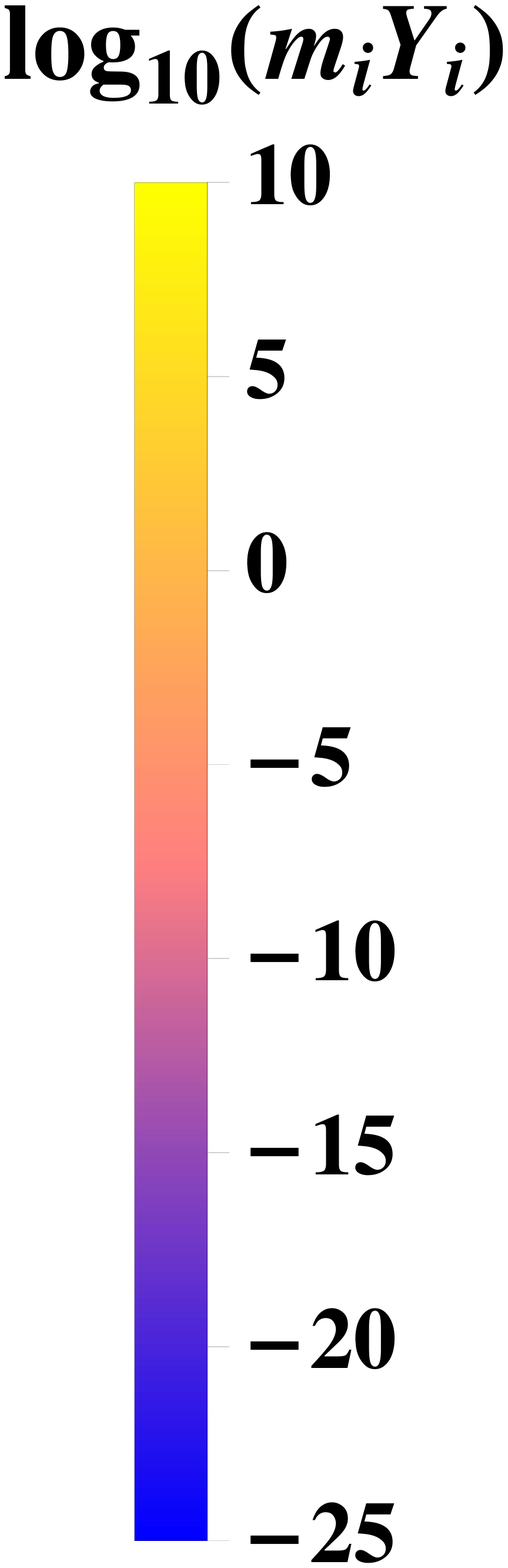}}
  \caption{Mass-weighted relic yields of the $\zpp$~(left) and $\opm$~(right) 
    glueballs in the $m_0$--$R$ plane in the absence of connectors 
    for $G_x=SU(3)$.  The solid white lines in each panel 
    indicate where the relic density saturates the observed 
    dark matter abundance.  The dark masked region at the lower right 
    of both panels shows where $\zpp$ freezeout occurs for $x_x^{fo} < 5$
    and our freezeout calculation is not applicable due to the unknown
    dynamics of the confining phase transition. 
    \label{fig:reld}}
\end{figure}

  In Fig.~\ref{fig:reld} we show the relic yields of $\zpp$~(left) 
and $\opm$~(right) glueballs in the absence of connectors to the SM 
in the $m_0$--$R$ plane for $G_x = SU(3)$.
The white lines in both panels indicate where the relic density of
that species coincides with the observed DM density,
$\Omega_{DM}h^2 = 0.1188(10)$~\cite{Ade:2015xua}.
The shaded regions at the lower right of both panels show where
$x_x^{fo} < 5$ implying the glueball densities are set by
the non-perturbative dynamics of the confining phase transition.
As expected, the $\opm$ yield is always much lower than the $\zpp$ yield.

  Going beyond the two-state model, our arguments regarding the exponential
suppression of the $\opm$ density relative to the $\zpp$ also apply to
the other heavier glueball modes~\cite{Forestell:2016qhc}.  
The total glueball relic density is strongly dominated by the 
$\zpp$ density, while $2\to 2$ annihilation reactions push the heavier 
glueball densities to much smaller values.  In fact, these reactions tend 
to be much more efficient for the other heavier glueballs than 
the $\opm$ due to coannihilation with the $\zpp$.
For example, $2^{++}+\zpp\to \zpp+\zpp$ efficiently depletes
the second-lightest $2^{++}$ glueball up to very large $x_x$,
while reactions such as $1^{--}+\zpp \to \opm+\zpp$ quickly transfer
the density of heavier $C$-odd glueballs to the lighter $\opm$.  
Conservation of $C_x$ number in the dark sector implies that the net density
of $C$-odd glueballs cannot be reduced by coannihilation.
As a result, the $\opm$ state generally develops the second largest relic 
density, with the densities of the other dark glueballs being much smaller.
This, combined with the unique decay properties of the $\opm$ glueball
when connectors are included, is the reason why we only consider
the effects of the $\zpp$ and $\opm$ glueballs in our analysis
of glueball cosmology.

\subsection{Glueball Freezeout with Connectors\label{sec:therm}}

Connector operators can modify the freezeout of glueballs in a number of ways.  
Scattering and decay reactions mediated by such operators transfer 
energy between the visible and dark sectors, and may allow them to thermalize.
Decays through the connector operators after confinement also deplete 
glueballs, and can occur before or after the freezeout of the various
$(3\to 2)$ and $(2\to 2)$ reactions.  We investigate these effects 
here, both before and after confinement, with a focus on the $\zpp$ 
and $\opm$ glueballs.  Our goal is to compute the yields of these species 
prior to their decay.

As in the freezeout analysis without connectors, we take as an initial condition
primordial inflation (or something like it) with preferential reheating 
to the visible sector characterized by a temperature $T_{RH}$ that 
is larger than the confinement transition temperature $T_c \simeq m_0/5.5$.  
With connectors, we also assume $T_{RH} \ll M$.  Reheating above the connector 
scale $M$ is likely to thermalize the dark and visible sectors at $T_{RH}$, 
and can produce a relic abundance of the connector particles themselves.  
These can have interesting cosmological effects in their own right, 
acting as \emph{quirks} if they carry $G_x$ charge~\cite{Okun:1980kw,Okun:2006eb,Burdman:2006tz},
and potentially creating dark glueballs non-thermally~\cite{Kang:2006yd,Jacoby:2007nw,Soni:2017nlm,Mitridate:2017oky}.
By taking $T_{RH} \ll M$, the production of connector particles in the 
early universe is strongly suppressed allowing us to focus on the effects
of the glueballs.

\subsubsection{Energy Transfer before Confinement}

  Consider first the transfer of energy at temperatures $T$ well
above the confinement temperature $T_c$.  In the absence of connectors, 
preferential reheating to the visible sector produces $T_x \ll T$.
Connector operators allow reactions of the form 
$\text{SM}+\text{SM}\leftrightarrow X+X$ that transfer energy from 
the visible sector to the dark sector.  For $T_x > T_c$, 
the evolution equation for the energy density of the dark sector
is~\cite{Bertschinger:2006nq,Bringmann:2006mu}
\beq
\frac{d\rho_x}{dt} + 4H\rho_x &=& -\langle\Delta E\ccdot \sigma v\rangle
\left(n_x^2-\widetilde{n}_x^2\right) \ ,
\label{eq:trans1}
\eeq   
where $\langle\Delta E\cdot\sigma v\rangle$ is the thermally averaged
energy transfer cross section for $X+X \to \text{SM}+\text{SM}$, $n_x$ is
the dark gluon number density, and 
$\widetilde{n}_x = \tilde{g}_x(\zeta(3)/\pi^2)T^3$ 
is the value it would have in full equilibrium with the visible sector 
with $\tilde{g}_x$ dark gluon degrees of freedom 
(equal to $\tilde{g}_x = 2(N^2\!-\!1)$ for $G_x=SU(N)$).\footnote{
Implicit in Eq.~\eqref{eq:trans1} is the assumption of self-thermalization 
of the energy injected into the dark sector to a temperature $T_x > T_c$.
Thermalization of non-Abelian gauge theories tends to be 
efficient~\cite{Kurkela:2011ti}, and we expect this assumption to 
be valid provided the total energy transfer is not exceedingly small.}
For $T_x \ll T$, the $\widetilde{n}_x^2$ term on the right side above
dominates and leads to a net energy transfer to the dark sector.
This transfer saturates and ceases when $T_x \to T$ 
and $n_x \to \widetilde{n}_x$.

For visible radiation domination with constant $g_*$, 
Eq.~\eqref{eq:trans1} can be rewritten as
\beq
\frac{d}{dT}\lrf{\rho_x}{T^4} = 
\frac{1}{HT^5}\langle\Delta E\ccdot\sigma v\rangle
\left(n_x^2-\widetilde{n}_x^2\right) \ .
\label{eq:trans2}
\eeq  
With the connector operators of Eqs.~(\ref{eq:dim8a},\ref{eq:dim6}) 
and $T\gg T_x$, the right side of Eq.~\eqref{eq:trans2} takes the
parametric form
\beq
\Delta\mathcal{C} &\equiv& \langle\Delta E\ccdot\sigma v\rangle
\left(n_x^2-\widetilde{n}_x^2\right)
\label{eq:delc1}
\\
&\sim& -D_n\frac{\mpl\,T^{n-2}}{M^n} \ ,
\label{eq:delc2}
\eeq
where $n = 4,\,8$.  Integrating from temperature $T$ to the 
reheating temperature $T_{RH}$, the approximate solution is
\beq
\lrf{\rho_x}{T^4} - \lrf{\rho_x}{T^4}_{RH} ~\sim~ 
\frac{D_n}{(n-1)}
\frac{\mpl T_{RH}^{n-1}}{M^n}
\left[1-\lrf{T}{T_{RH}}^{n-1}\right] \ ,
\label{eq:delrhox1}
\eeq
This expression is dominated by the contribution
near the reheating temperature, and  represents the contribution 
to the dark energy density from transfer reactions.  

The approximate forms of Eqs.~(\ref{eq:delc2},\ref{eq:delrhox1}) 
are only valid for $T < T_{RH}$ and $T > T_x \geq T_c$.
The first of these conditions corresponds to the upper limit on the era
of radiation domination.  An even higher radiation temperature can be
achieved prior to reheating, but for standard perturbative reheating
and $n < 29/3 \simeq 9.67$ we find that the energy transfer before
the radiation era is also dominated by reactions near $T\sim T_{RH}$.  
The second condition $T > T_x \geq T_c$
is needed to justify our neglect of the $n_x^2$ term on the right side of 
Eq.~\eqref{eq:trans2} and our assumption of a deconfined phase.  
As $T_x$ approaches $T$ due to the energy transfer, 
this term becomes important and the net energy transfer goes to zero,
corresponding to the thermalization of the two sectors.  
 
  Motivated by these considerations, let us define
\beq
\Delta\!\lrf{\rho_x}{T^4} &\equiv& \int_\cdot^{T}\!dT'\;
\lrf{\Delta\mathcal{C}}{H{T'}^5}
\label{eq:delrhox2a}\\
&\sim& \frac{D_n}{(n-1)}
\frac{\mpl T^{n-1}}{M^n} \ .
\label{eq:delrhox2b}
\eeq
This represents the contribution to the dark sector
energy from thermal transfer in the vicinity of temperature $T$.
Thermalization occurs when
\beq
\Delta\!\lrf{\rho_x}{T^4} ~\geq~ \frac{\pi^2}{30}\tilde{g}_x \ ,
\label{eq:therm1}
\eeq
where $\tilde{g}_x$ is the number of dark gluon degrees of freedom.
Let $T_{th}$ be the temperature that solves Eq.~\eqref{eq:therm1} 
as an equality.  If $T_{th} < T_c$, the visible and dark sectors 
remain thermalized at least until confinement.  
Conversely, if $T_{th} > T_c$ thermalization
is lost at $T=T_{th}$ and the dark and visible sectors evolve
independently thereafter with separately conserved entropies.

  The dark to visible entropy ratio $R$ is constant for $T < T_{th}$
and depends on reheating.  If $T_{th} < T_{RH}$, thermalization occurs 
after reheating and is maintained until $T = T_{th}$.
The entropy ratio $R$ (for $T_{th} > T_c$) after thermalization ceases 
is then 
\beq
R = R_{max} \equiv \frac{\tilde{g}_{x}}{g_{*S}(T_{th})} \ .
\label{eq:rmax}
\eeq
Thermalization need never have occurred after reheating if $T_{RH} < T_{th}$.
In this case, (for $T_{th} > T_c$) we can define
\beq
T_{x\,RH} = 
T_{RH}\left[\frac{30}{\pi^2\tilde{g}_x}\;\Delta\!\lrf{\rho_x}{T^4}_{RH}\right]^{1/4}
\ .
\eeq
This implies a lower bound on the entropy ratio of
\beq
R \geq \frac{\tilde{g}_{x}}{g_{*S}(T_{RH})}\lrf{T_{x\,RH}}{T_{RH}}^3 \ .
\label{eq:rmin}
\eeq
In general, lower reheating temperatures allow for smaller
values of $R$.  We define $R_{min}$ to be the value of $R$
such that $T_{x\,RH} = T_c$, the lowest possible reheating temperature
given our assumption of $T_{x\,RH} \geq T_c$.\footnote{
Even lower values of $R$ are possible for $T_{x\,RH} < T_c$, but this also
implies that reheating can interfere with the freezeout 
process~\cite{Moroi:1999zb}, and goes beyond the scope of this work.}
When $T_{th} > T_c$, the range of $R$ values is therefore
$R_{min} \leq R \leq R_{max}$.  

\begin{figure}[ttt]
  \centering
	\includegraphics[width=0.35\textwidth]{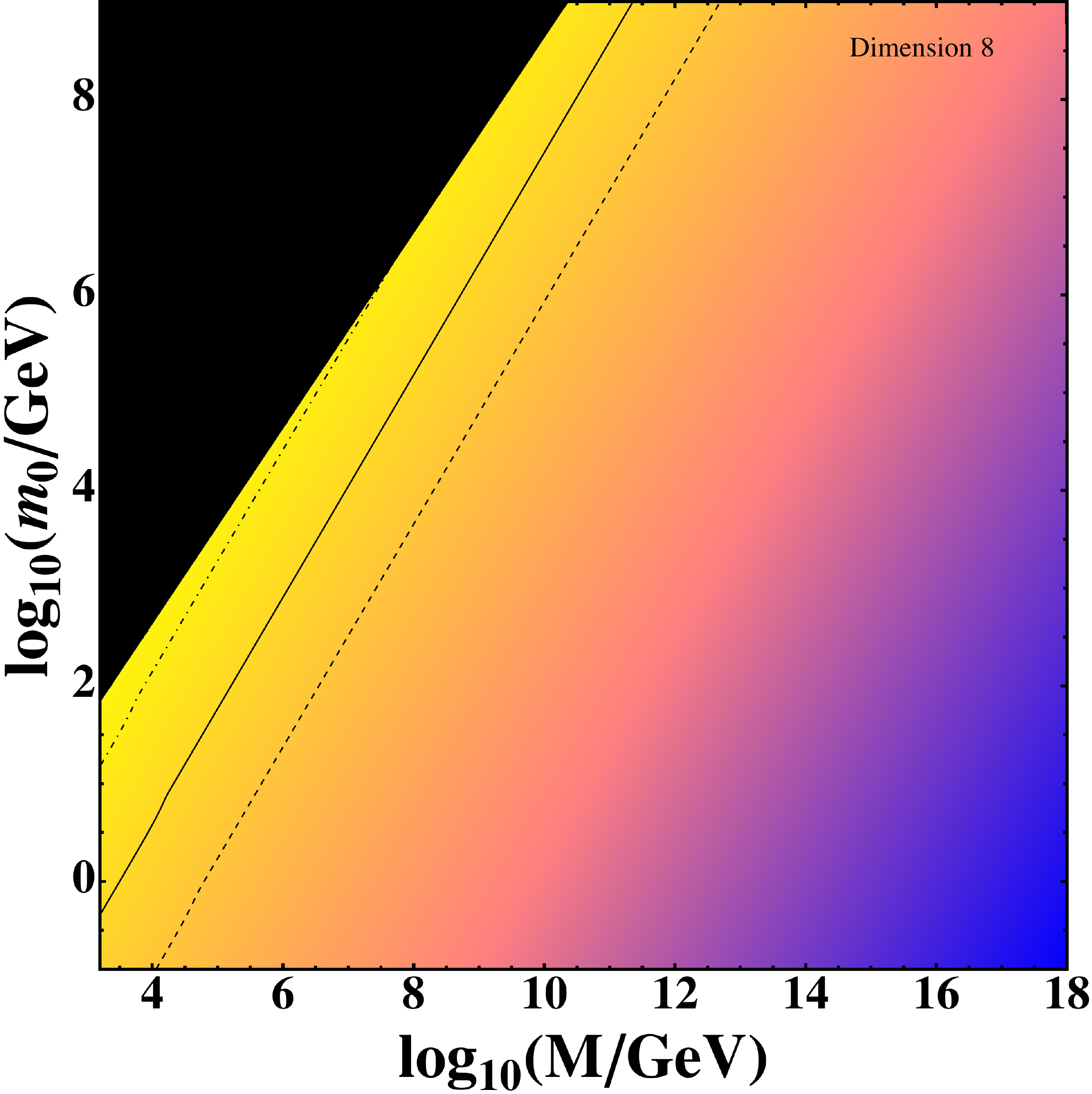}
	\includegraphics[width=0.35\textwidth]{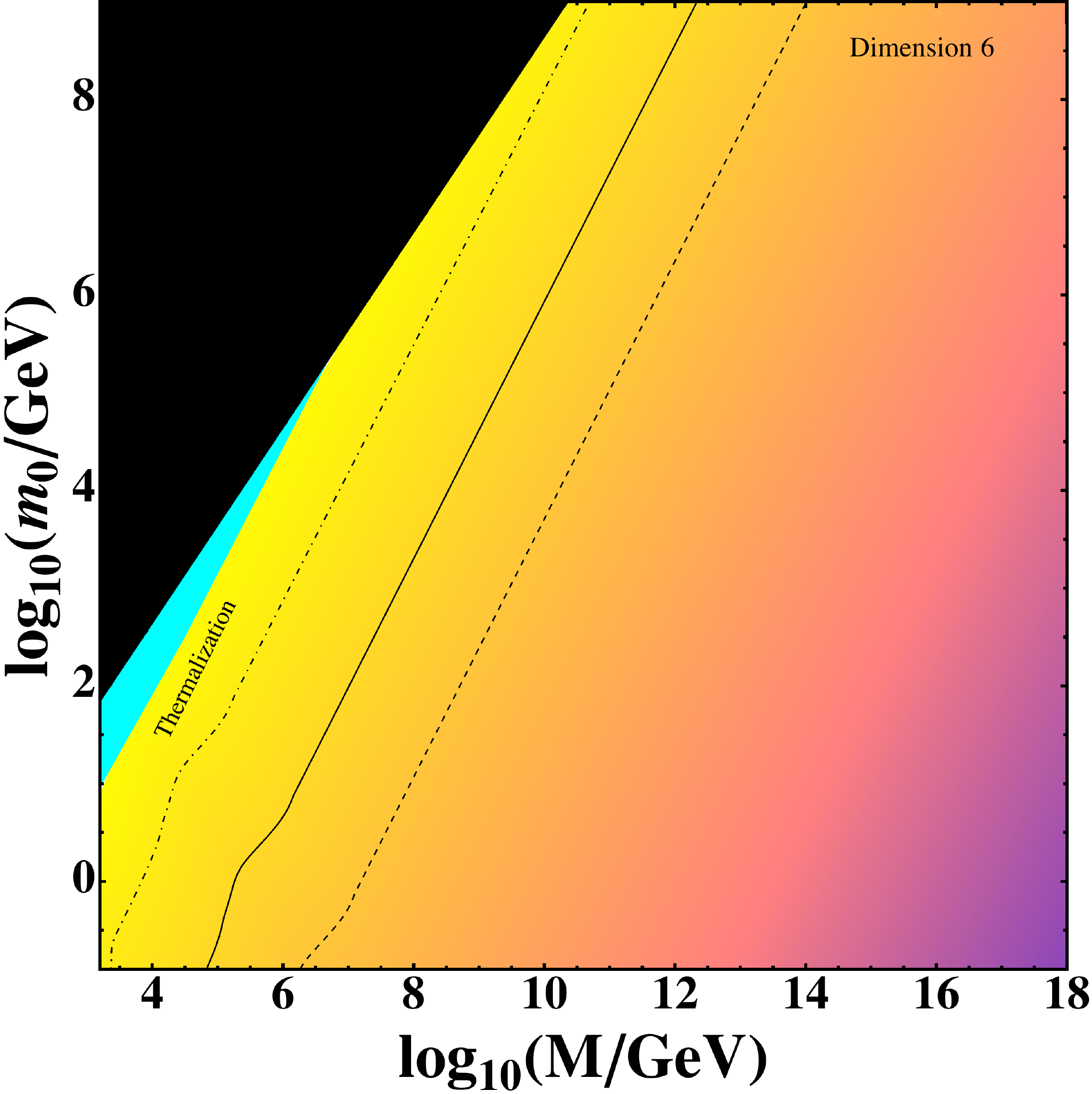}
	\raisebox{1.0cm}{\includegraphics[width=0.2\textwidth]{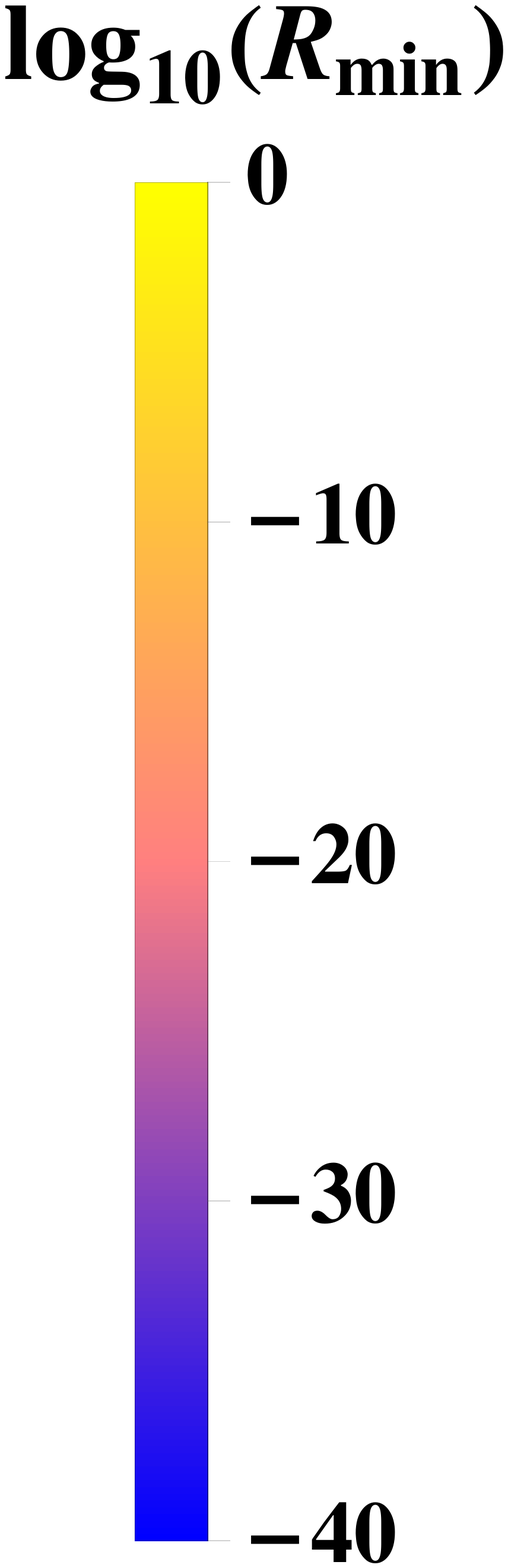}}
  \caption{Values of the minimal entropy ratio $R_{min}$ in the $M$--$m_0$
plane for energy transfer via dimension-8~(left) and dimension-6~(right) 
operators for $G_x=SU(3)$.  The black shaded region at the upper left indicates
where our treatment in terms of effective operators breaks down. 
The diagonal black dotted, solid, and dashed lines show reference values 
of $R_{min} = 10^{-3},\,10^{-6},\,10^{-9}$.  In the cyan region in the right panel,
thermalization between the visible and dark sectors is maintained 
at least until confinement.
    \label{fig:rmin}}
\end{figure}

  In Appendix~\ref{sec:appa} we present explicit expressions for
the collision term $\Delta\mathcal{C}$ needed to compute the 
energy transfer $\Delta(\rho_x/T^4)$ via Eq.~\eqref{eq:delrhox2a}.
The results obtained for $R_{min}$ are shown in Fig.~\ref{fig:rmin} 
in the $m_0$--$M$ plane for energy transfer via dimension-8~(left) 
and dimension-6~(right) operators for $G_x=SU(3)$.  The shaded region 
at the upper left has $m_0 > M/10$ and indicates where our treatment
in terms of effective operators breaks down.  The black dotted, solid, and 
dashed lines show reference values of $R_{min} = 10^{-3},\,10^{-6},\,10^{-9}$.
In the cyan region in the right panel, thermalization between the visible
and dark sectors is maintained at least until confinement, corresponding
to $T_{th} < T_c$.

\subsubsection{Evolution of the $\zpp$ Density}

  Glueballs form at $T_x=T_c$ and undergo freezeout, transfer, 
and decay reactions.  In the absence of connectors, the dominant glueball
species is the lightest $\zpp$ mode.  To track its evolution with
connector operators, it is convenient to organize the analysis according
to the thermalization temperature $T_{th}$, computed above 
in the unconfined phase, relative to the confinement temperature.

\bigskip 

\noindent
\textbf{$\mathbf{T_{th} < T_c}$:}~
This condition implies that thermalization is maintained at least 
until confinement, and thus we expect $T=T_x=T_c$ as an initial
condition for the glueball evolution.  To compute the $\zpp$ density 
and thermal transfer after confinement we adapt the analysis of 
Refs.~\cite{Kuflik:2015isi,Kuflik:2017iqs} 
based on Refs.~\cite{Bertschinger:2006nq,Bringmann:2006mu,Gondolo:2012vh},
which is applicable here since $T,\,\,T_x \leq T_c \simeq m_0/5.5$.  
If thermal equilibrium is maintained independently within both 
the dark and visible sectors,
the dark temperature evolves as~\cite{Kuflik:2015isi,Kuflik:2017iqs} 
\beq
\frac{dT_x}{dt} ~\simeq~ -2HT_x
+ \frac{2}{3n_0}(\mathcal{C}_{\rho}-m_0\,\mathcal{C}_n)
\label{eq:dtxdt}
\eeq
where $\mathcal{C}_{\rho}$ and $\mathcal{C}_n$ are the collision
terms appearing in the evolution equations for the $\zpp$ energy 
and number densities.  The Hubble term in Eq.~\eqref{eq:dtxdt} gives 
the usual $1/a^{2}$ redshifting of the effective temperature of 
an independent massive species, while the second term describes 
energy transfer from scattering and decay processes.

  The explicit forms of the collision terms are
\beq
\mathcal{C}_n ~\simeq~ -\langle\sigma_{32}v^2\rangle n_0^2(n_0-\bar{n}_0)
-\Gamma_0\left[n_0(1-3T_x/2m_0) - \widetilde{n}_0(1-3T/2m_0)\right]
\ ,
\label{eq:cn}
\eeq
where $\bar{n}_0 = n_0(T_x)$ and $\widetilde{n}_0 = n_0(T)$,
as well as
\beq
\mathcal{C}_{\rho} ~\simeq~ 
n_0n_{SM}\langle\sigma_{el}v\ccdot\Delta E\rangle
- m_0\Gamma_0(n_0-\widetilde{n}_0) \ .
\label{eq:cp}
\eeq
The only new piece in these expressions is the elastic scattering term
$n_0n_{SM}\langle\sigma_{el}v\ccdot\Delta E\rangle$ in Eq.~\eqref{eq:cp}.
It corresponds to reactions of the form $\text{SM}+\zpp \to \text{SM}+\zpp$,
and was studied in detail in Refs.~\cite{Bertschinger:2006nq,Bringmann:2006mu}.  

Combined in Eq.~\eqref{eq:dtxdt}, the $(3\to 2)$ scattering term 
from Eq.~\eqref{eq:cn} tends to heat the dark glueballs, 
and the elastic scattering and decay terms tend to drive $T_x\to T$.
Applied to the $\zpp$ glueball with either the dimension-8 or dimension-6 
connector operators, we find that thermalization below confinement 
implies $\Gamma_0 > H(T=m_0)$.  Thus, the $\zpp$ density simply
tracks the equilibrium value with temperature $T$ following confinement.\footnote{
In the absence of decays, massive glueballs with connectors would give
an explicit realization of the SIMP~\cite{Hochberg:2014dra} 
or ELDER~\cite{Kuflik:2015isi,Kuflik:2017iqs} DM scenarios.}
  
\bigskip

\noindent
\textbf{$\mathbf{T_{th} > T_c}$:}~
With $T_{th} > T_c$, the visible and dark sectors are not thermally
connected at confinement, and thus $T \geq T_x$ at this point 
with a well-defined entropy ratio in the range $R_{min} \leq R \leq R_{max}$.  
Using the scaling arguments applied above, it can be shown that 
$R \geq R_{min}$ implies $T \leq m_0$ 
when the $\zpp$ decays set in at $\Gamma_0 \simeq H(T)$.\footnote{
Our numerical analysis confirms this as well.}  The evolution equations
for the $\zpp$ number density and temperature can thus be written as
(to leading order in $T_x/m_0$)
\beq
\frac{dn_0}{dt} &=& -3Hn_0 -\langle\sigma_{32}v^2\rangle n_0^2(n_0-\bar{n}_0)
-\Gamma_0(n_0-\widetilde{n}_0) 
\label{eq:subtc1}\\
&&\nnmb\\
\frac{dT_x}{dt} &=& -2HT_x 
+ \frac{2}{3}m_0\langle\sigma_{32}v^2\rangle n_0(n_0-\bar{n}_0)
+ \Gamma_0T_x\left(1-\frac{\widetilde{n}_0}{n_0}\frac{T}{T_x}\right)
\label{eq:subtc2}
\eeq  
where again $\bar{n}_0$ is the equilibrium value at temperature $T_x$
and $\widetilde{n}_0$ is the equilibrium value at temperature $T$.
Note that we have neglected the elastic scattering term because
it can be shown to be parametrically small relative to the Hubble 
term for $T< T_{th}$ and $R\geq R_{min}$.  

  When the decay terms are neglected, the evolution equations 
of Eqs.~(\ref{eq:subtc1},\,\ref{eq:subtc2}) are equivalent 
to those we used previously with no connector operators 
(to leading order in $T_x/m_0$).
Glueball decays only become significant when $\Gamma_0 \simeq H(T)$, 
and quickly drive $T_x\to T$ and $n_0\to \widetilde{n}_0$. 
It follows that our previous analysis without connectors can be applied 
to compute the $\zpp$ relic yield prior to decay (which may occur
before freezeout).  
The only significant effect of energy transfer on this calculation 
is to limit the range of the initial entropy ratio to 
$R_{min}\leq R \leq R_{max}$.

\subsubsection{Evolution of the $\opm$ Density\label{sec:ydecay}}

  Even though the lightest $\zpp$ glueball dominates the total glueball 
density and controls the dark temperature prior (and even after) its decay,
the heavier $\opm$ glueball can also be relevant for cosmology due
to its longer lifetime.
Recall that the $\opm$ is parametrically long-lived
relative the $\zpp$ in the decay scenarios~\textbf{2}--\textbf{4} 
listed in Sec.~\ref{sec:dscenario}, where the $\zpp$ decays through 
a dimension-6 operator while the $\opm$ is stable or only decays 
at dimension-8.  Even in decay scenario~\textbf{1}, where both states 
decay at dimension-8, the $\zpp$ decay rate tends to be larger than 
the $\opm$ by a factor of $(N_c^2-1)\alpha_3/\alpha$.  

  The evolution of the $\opm$ density is sensitive to the $\zpp$ density
in several ways.  Prior to decay, the $\zpp$ density acts as a massive
thermal bath that cools very slowly relative to the visible temperature,
thereby delaying the freezeout of the $\opm$ state.
This thermal bath collapses and disappears
when the $\zpp$ decays, which can hasten $\opm$ freezeout.
If the $\zpp$ density is large when it decays, the entropy transferred
to the visible sector can also dilute the densities of the remaining
$\opm$ glueballs.  We investigate these effects here, 
dividing the analysis into $T_{th} < T_c$ and $T_{th} > T_c$ cases.

\bigskip
\noindent
\textbf{$\mathbf{T_{th} < T_c}$:}~
Recall that this case is only realized for dimension-6
transfer operators, and implies that the $\zpp$ decay rate is larger
than Hubble following confinement.  This means the $\zpp$
density tracks its equilibrium value with effective temperature $T_x=T$,
and there is no longer a separately conserved entropy in the dark sector.
The evolution of the $\opm$ number density in this context is
\beq
\frac{dn_1}{dt} +3Hn_1 =
-\langle\sigma_{22} v\rangle\left(n_1^2-\widetilde{n}_1^2\right) 
- \Gamma_1(n_1-\widetilde{n}_1)
\ ,
\label{eq:evol1therm}
\eeq
where $\widetilde{n}_1$ denotes the equilibrium density of the $\opm$
at temperature $T$.  Note that Eq.~\eqref{eq:evol1therm} assumes
the $\opm$ mode also thermalizes with the visible sector.  This is expected
prior to freezeout since the equilibrium density of the $\opm$ is smaller
than that of the $\zpp$, and elastic scattering between these two
species is at least as efficient as the annihilation reaction.

\bigskip
\noindent
\textbf{$\mathbf{T_{th} > T_c}$:}~
This second case implies $T_x \leq T$ at confinement, with $\zpp$ decays
inactive ($\Gamma_0 < H$) until $T < m_0$.  To compute the resulting
$\opm$ relic density, we treat the $\zpp$ decay as instantaneous
and match the density evolution immediately before and after it occurs.
Prior to the decay, the dark and visible entropies are
conserved independently with ratio $R$, and the glueball densities
evolve according to Eqs.~(\ref{eq:evol0},\ref{eq:evol1}). 
Decays of the $\zpp$ are implemented at $\Gamma_0 = H$,
where the Hubble rate includes contributions from both the visible
and dark energy densities.  If $T_{i} < m_0$ is the visible temperature
prior to the decay, the visible temperature afterwards is obtained
from local energy conservation,
\beq
\rho(T_f) =  \rho(T_i) + \rho_x(T_i)  \ ,
\eeq
where we have neglected the exponentially subleading contribution of the $\opm$
mode to the energy density.  Note that $T_f > T_i$ is always smaller than $m_0$
as well.  The evolution of the $\opm$ number density after the $\zpp$ decays
is given by Eq.~\eqref{eq:evol1therm}.  Since the 
$\opm$ number density is not changed by the decays,
$n_1(T_f) = n_1(T_i)$ is used as the initial condition at $T=T_f$.

  The interplay of glueball annihilation, transfer, and decays
leads to many different qualitative behaviours.  These were investigated
in Ref.~\cite{Pappadopulo:2016pkp,Farina:2016llk} for a simplified model 
consisting of an unstable massive bath particle and a heavier DM state.  
Dark glueballs provide an explicit realization of this scenario, 
with the $\zpp$ making up the massive bath and the $\opm$ acting 
as (metastable) dark matter.  Compared to the simple model studied
in Ref.~\cite{Pappadopulo:2016pkp,Farina:2016llk}, the $\zpp$ massive 
bath particle always freezes out (or decays) before the would-be
$\opm$ dark matter, corresponding to the \emph{chemical} or \emph{decay}
scenarios discussed there.  A potential further behavior that we have
not captured in our approximations is the late production of $\opm$
glueballs through transfer reactions while $T > m_1$ but after $\opm$ 
freezeout has occurred in the dark sector.  We estimate that this is
potentially relevant in a very limited corner of the parameter space,
and will only increase the limits we find.


\subsection{Comments on Theoretical Uncertainties\label{sec:uncrt}}

  Before applying our results for dark glueball lifetimes and densities
to derive cosmological and astrophysical constraints on them, 
it is worth taking stock of the theoretical uncertainties in our calculations.  
It is also useful to identify how some of these uncertainties might be reduced 
with improved lattice calculations.  

  The glueball lifetimes computed in Sec.~\ref{sec:decay} rely on glueball
masses and transition matrix elements.  Masses for $G_x = SU(3)$ have
been obtained to a precision greater than 5\% 
in Refs.~\cite{Morningstar:1999rf,Chen:2005mg}, while the matrix element
relevant for $\zpp$ decay was determined to about 20\% in 
Refs.~\cite{Chen:2005mg,Meyer:2008tr}.  Thus, we expect our determination
of the $\zpp$ decay width to be reasonably accurate.  The situation is
less clear for the $\opm$ width, which relies on a $\opm \to \zpp$
transition matrix element that we were only able to estimate using NDA.
In the absence of lattice calculations for this matrix element, 
we estimate that our $\opm$ width is only reliable to within a factor of a few.

  Turning next to the cosmological evolution of the dark gluons and glueballs,
we implicitly treated their interactions as being perturbative.  This is
a good approximation at temperatures well above the confinement scale,
but significant deviations can arise as the temperature falls to near
confinement~\cite{Borsanyi:2012ve}.  For the range of entropy ratios 
$R$ due to energy transfer computed above, this implies that values of $R_{max}$
with $T_{th} \gg T_c$ are reliable, but the specific values of 
$R_{min}$ and $R_{max}$ for $T_{RH} \sim T_c$ could receive large
corrections.  Similarly, the glueball interactions used to compute 
the $(3\to 2)$ and $(2\to 2)$ cross sections are quite strong for $N = 3$.
It is difficult to quantify how this affects the pre-decay
glueball relic densities, but we do note that the densities typically depend 
roughly linearly on $R$ and the annihilation cross sections.
Our naive estimate is that the pre-decay glueball densities we find 
are accurate to within about an order of magnitude.

\section{Cosmological Constraints\label{sec:constraint}}

In the analysis above we showed that dark glueballs can have a wide
range of decay rates and a variety of formation histories in the early
universe.  Very long-lived dark glueballs can potentially make up the 
cosmological dark matter.  On the other hand, shorter-lived glueballs are
strongly constrained by the modifications they can induce in
the standard predictions for big bang 
nucleosynthesis~(BBN)~\cite{Kawasaki:2004qu,Jedamzik:2006xz,Kawasaki:2017bqm}, 
the cosmic microwave background~(CMB)~\cite{Chen:2003gz,Slatyer:2016qyl}
and the spectrum of cosmic rays~\cite{Cohen:2016uyg}.
We investigate the bounds from cosmology and astrophysics
on dark glueballs in this section for the four decay
scenarios discussed in Sec.~\ref{sec:decay}.  Throughout the analysis,
we focus on $G_x = SU(N\!=\!3)$, and we assume reheating such that $T_{RH} \ll M$
and $T_{x\,RH}\geq T_c$.  Details of how we implement the bounds 
from BBN, the CMB, and cosmic rays are collected in Appendix~\ref{sec:appb}.

\subsection{Decay Scenario 1: Dimension-8 Decays with Broken $C_x$}

\begin{figure}[ttt]
	\centering
${}$\hspace{0.5cm}
	\includegraphics[width=0.13\textwidth]{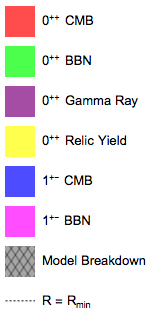}
\hspace{1.2cm}
	\includegraphics[width=0.25\textwidth]{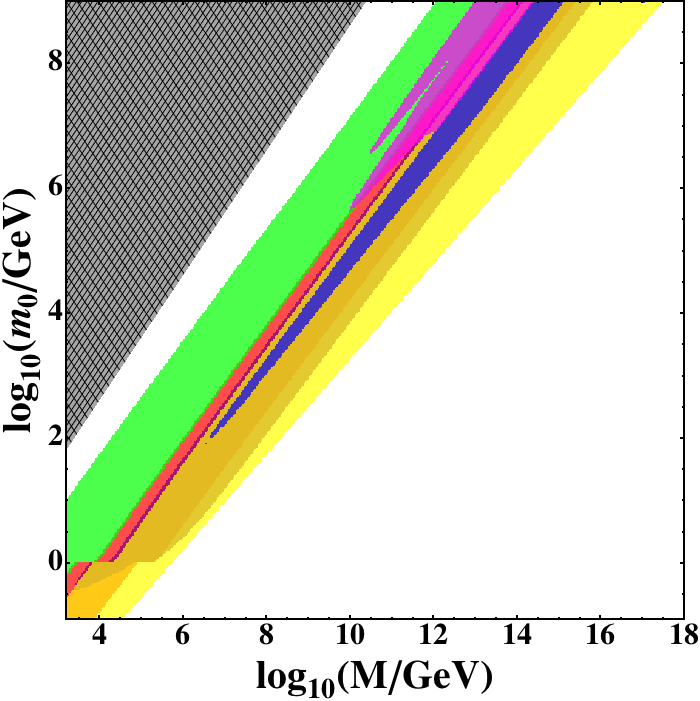}
	\includegraphics[width=0.25\textwidth]{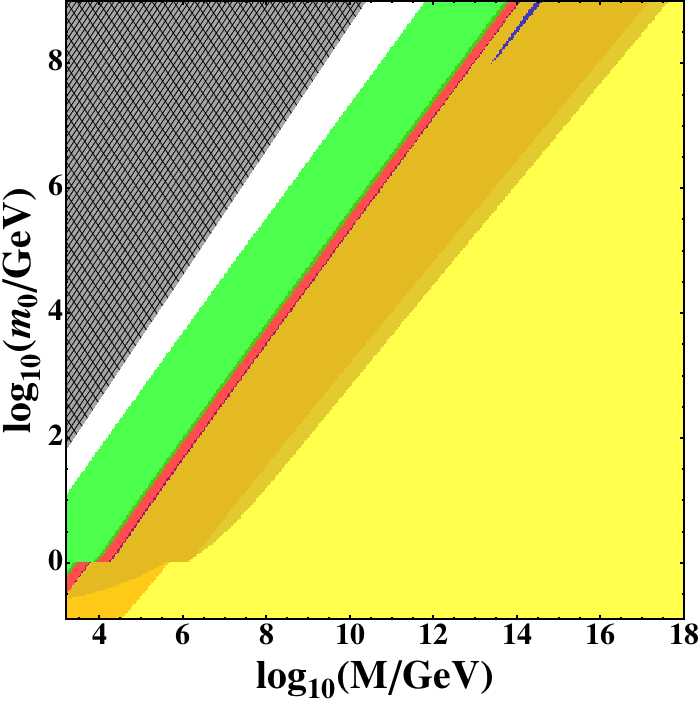}
\vspace{0.5cm}\\
	\includegraphics[width=0.25\textwidth]{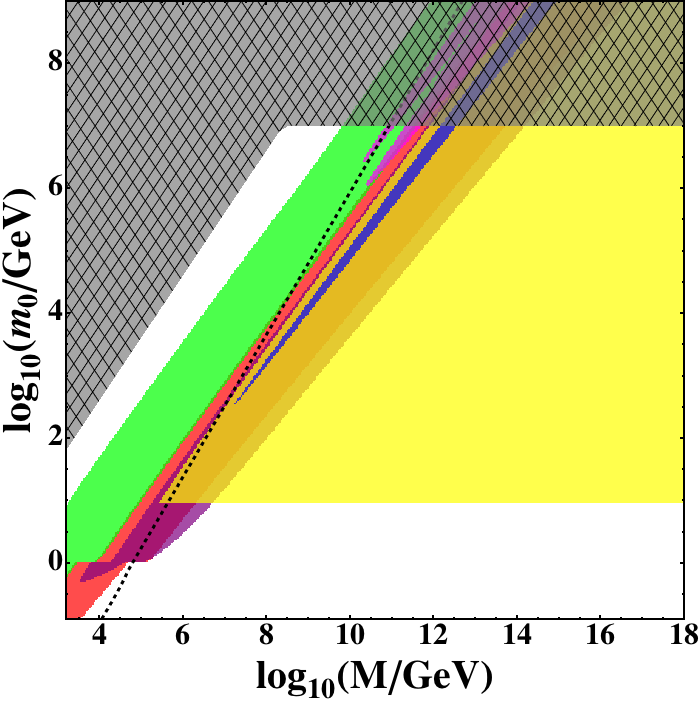}
	\includegraphics[width=0.25\textwidth]{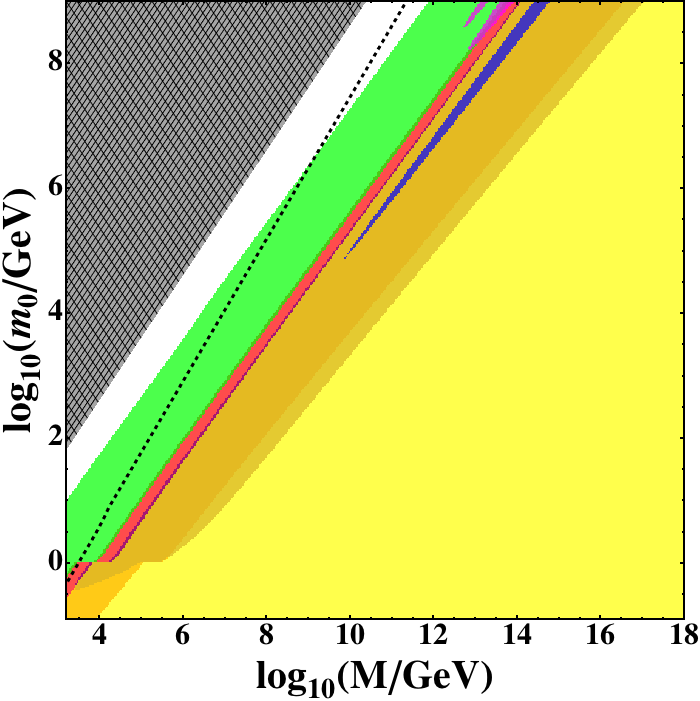}
	\includegraphics[width=0.25\textwidth]{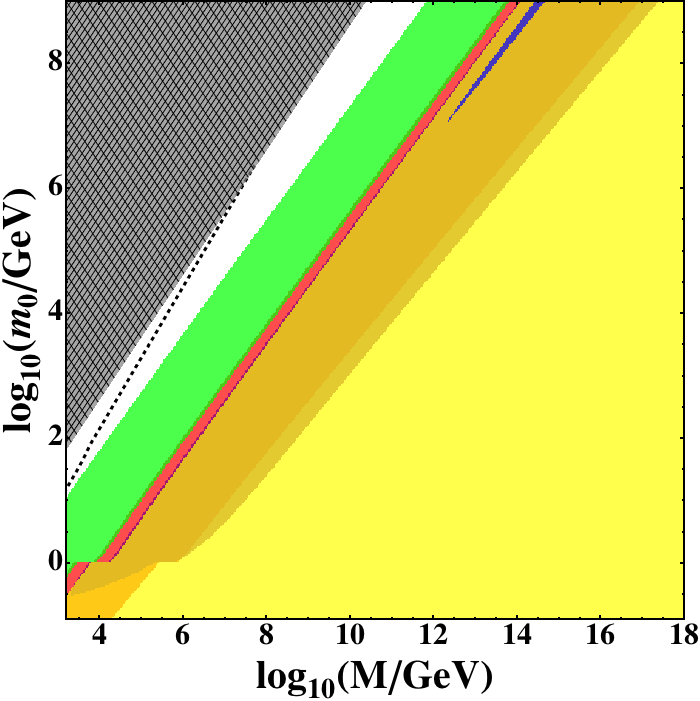}
	\vspace{-0.1cm}
	\caption{Cosmological constraints on dark glueballs 
in the $M$--$m_0$ plane for decay scenario~\textbf{1}
with dominant dimension-8 operators and broken $C_x$.  The upper two
panels have $R=R_{min},\,R_{max}$, while the lower three panels have
fixed $R=10^{-9},\,10^{-6},\,10^{-3}$.  The grey shaded region in each
panel indicates where our theoretical assumptions fail, 
while $R < R_{min}$ to the left of the dashed line.  
\label{fig:dec_s1}}
\end{figure}

   This scenario has all the dimension-8 operators 
of Eq.\eqref{eq:leff8} with $\chi_i=\chi_Y = 1$.  
Both the $\zpp$ and $\opm$ glueballs decay with parametrically similar rates,
as shown in Fig.~\ref{fig:decrates8}.  

The cosmological constraints on this scenario are shown in Fig.~\ref{fig:dec_s1}
in the $M$-$m_0$ plane for various values of the entropy ratio $R$.  
The upper two panels have $R=R_{min},\,R_{max}$ respectively,\footnote{
Recall from Eq.~\eqref{eq:rmax} that $R_{max}$ corresponds to thermalization
after reheating, while from Eq.~\eqref{eq:rmin} $R_{min}$ is the lowest
possible entropy ratio consistent with energy transfer and $T_{x\,RH}>T_c$.}
and the lower three panels show $R=10^{-9},\,10^{-6},\,10^{-3}$.  
The grey shaded regions indicate where our theoretical
assumptions break down.  The rising diagonal portion of the gray shaded
region corresponds to $m_0 > M/10$; we demand smaller values of $m_0$
to justify our treatment in terms effective operators suppressed by powers of $M$.
The upper part of the grey shaded region indicates $T_{x\,fo} > T_c$,
corresponding to glueball densities set by the non-perturbative dynamics
of the confining phase transition.  To the left of the diagonal dotted lines
in the lower three panels, the given fixed value of $R$ is less than $R_{min}$
and is inconsistent with minimal energy transfer by the connector operators 
for $T_{x\,RH}> T_c$.

  We see from Fig.~\ref{fig:dec_s1} that dark glueballs are strongly 
constrained by cosmological and astrophysical observations.  
When the $\zpp$ is long-lived, corresponding to small $m_0/M$,
its relic density tends to be too large unless the entropy ratio $R$ 
is much less than unity.  With sufficiently small $R$
the $\zpp$ can make up all the dark matter corresponding to the white
line in the left panel of Fig.~\ref{fig:reld}.  Such a DM candidate
would be very difficult to probe, with the most promising avenues
being high energy gamma rays and modifications to cosmic structure
from glueball self interactions.  Using large-$N$ and NDA,
the $2\to 2$ self-interaction cross section of $\zpp$ glueballs
is~\cite{Boddy:2014yra,Soni:2016gzf}
\beq
\sigma_{2\to 2}/m_0 ~\simeq~ (10\,\text{cm}^2/\text{g})\lrf{3}{N}^4
\lrf{100\,\mev}{m_0}^3 \ .
\eeq
This is at (or slightly above) the current limit for $N \geq 3$ 
and $m_0 \geq 100\,\mev$ and could have observable effects close 
to these values~\cite{Tulin:2017ara}, but falls off very quickly 
with higher mass or if the $\zpp$ glueball is only a small fraction
of the full DM density. For larger $m_0/M$ ratios, 
the $\zpp$ and $\opm$ glueballs both decay 
quickly enough to alter BBN or the CMB or create high energy gamma rays.  
Not surprisingly, the bounds from glueball decays in this scenario 
come primarily from the $\zpp$ which has a much larger relic yield prior to decay.

\subsection{Decay Scenario 2: Dimension-8 Decays with Exact $C_x$}

\begin{figure}[ttt]
	\centering
${}$\hspace{0.8cm}
	\includegraphics[width=0.18\textwidth]{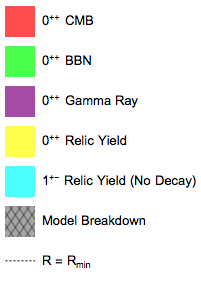}
\hspace{0.32cm}
	\includegraphics[width=0.25\textwidth]{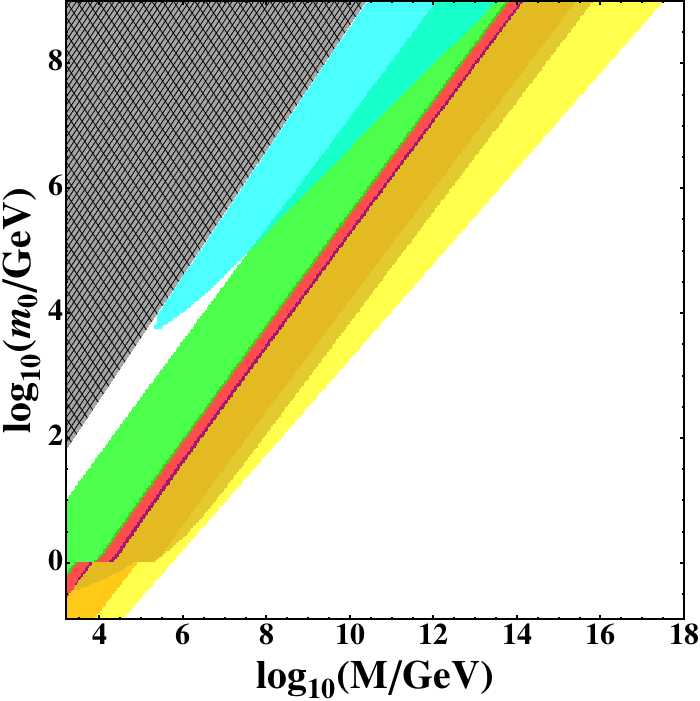}
	\includegraphics[width=0.25\textwidth]{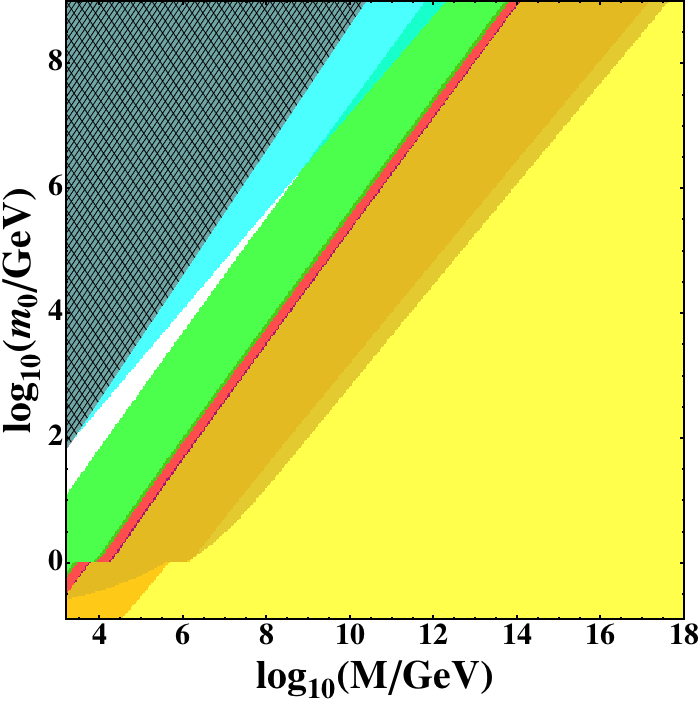}
\vspace{0.5cm}\\
	\includegraphics[width=0.25\textwidth]{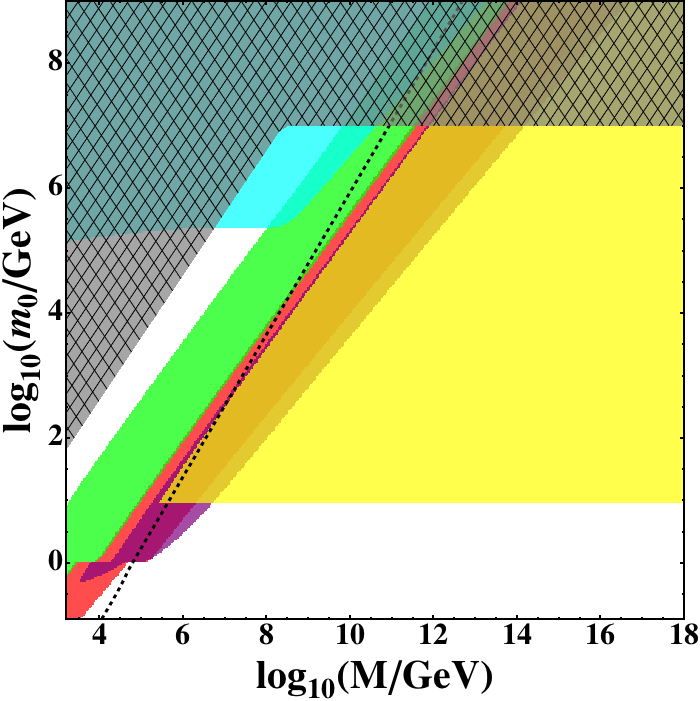}
	\includegraphics[width=0.25\textwidth]{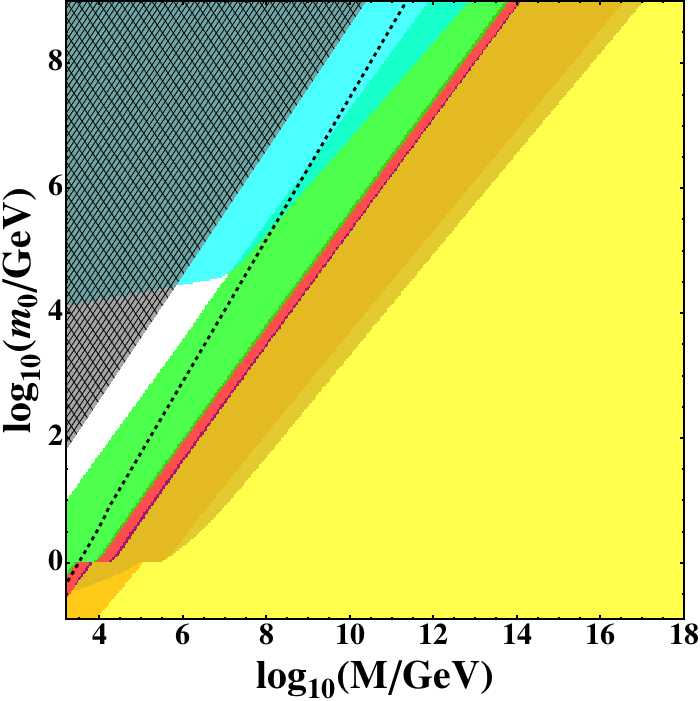}
	\includegraphics[width=0.25\textwidth]{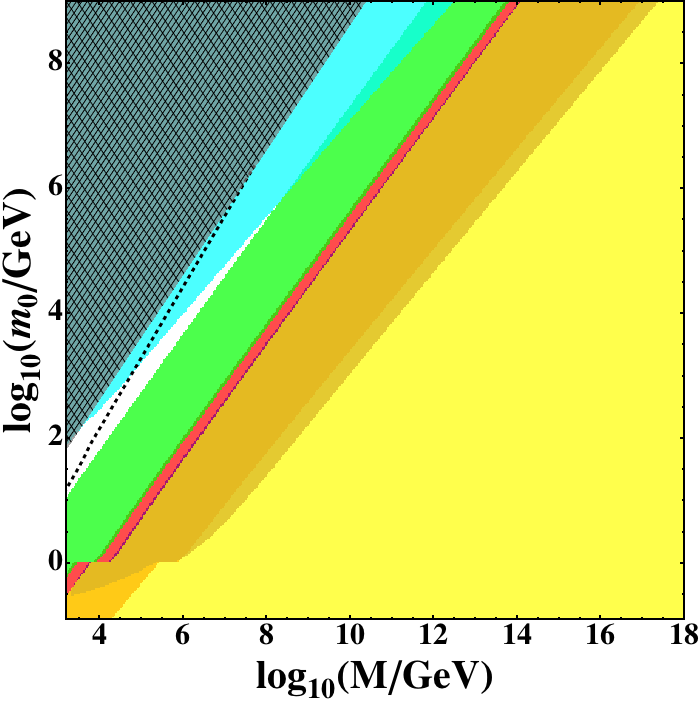}
	\vspace{-0.1cm}
	\caption{
Cosmological constraints on dark glueballs 
in the $M$--$m_0$ plane for decay scenario~\textbf{2}
with dominant dimension-8 operators and conserved $C_x$.  
The upper two panels have $R=R_{min},\,R_{max}$, and the lower three panels have
fixed $R=10^{-9},\,10^{-6},\,10^{-3}$.  The grey shaded region indicates where 
our theoretical assumptions fail, while to the left of the dashed line 
we find $R < R_{min}$. 
\label{fig:dec_s2}}
\end{figure}

  Our second decay scenario has dominant dimension-8 operators
with $\chi_i=1$ and a conserved $C_x$ charge that implies $\chi_Y=0$
and a stable $\opm$ glueball.  The cosmological bounds
on this scenario are shown in Fig.~\ref{fig:dec_s2}
for various values of the entropy ratio $R$.  
The upper two panels have $R=R_{min},\,R_{max}$ respectively,
and the lower three panels show $R=10^{-9},\,10^{-6},\,10^{-3}$.
As above, the grey shaded regions indicate where our theoretical
assumptions are not satisfied, and the diagonal dashed lines 
have $R< R_{min}$ to their left.

  The cosmological exclusions on this scenario are nearly identical to those
on scenario~\textbf{1} except for the new bounds from the $\opm$ relic density.
At the lower edge of the cyan excluded region, the $\opm$ glueball can make 
up all the dark matter.  This occurs primarily when the $\zpp$ decays relatively 
quickly, since otherwise it tends to dilute the $\opm$ relic density too strongly.
Note as well that the $\opm$ glueball can make up the dark matter for
a wide range of values of the entropy ratio $R$, and for masses well
above the weak scale, between about $10^2\,\gev \lesssim 10^5\,\gev$.
For smaller values of $m_0/M$, the $\zpp$ is long-lived and remains
the dominant species as in scenario~\textbf{1}.

\subsection{Decay Scenario 3: Dimension-6 Decays with Broken $C_x$}

\begin{figure}[ttt]
	\centering
${}$\hspace{0.5cm}
	\includegraphics[width=0.13\textwidth]{legend_decay}
\hspace{1.2cm}
	\includegraphics[width=0.25\textwidth]{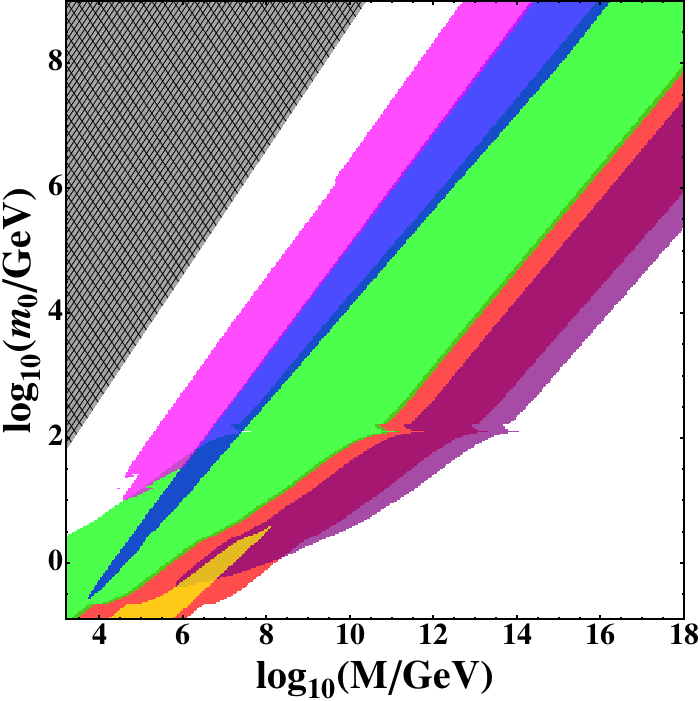}
	\includegraphics[width=0.25\textwidth]{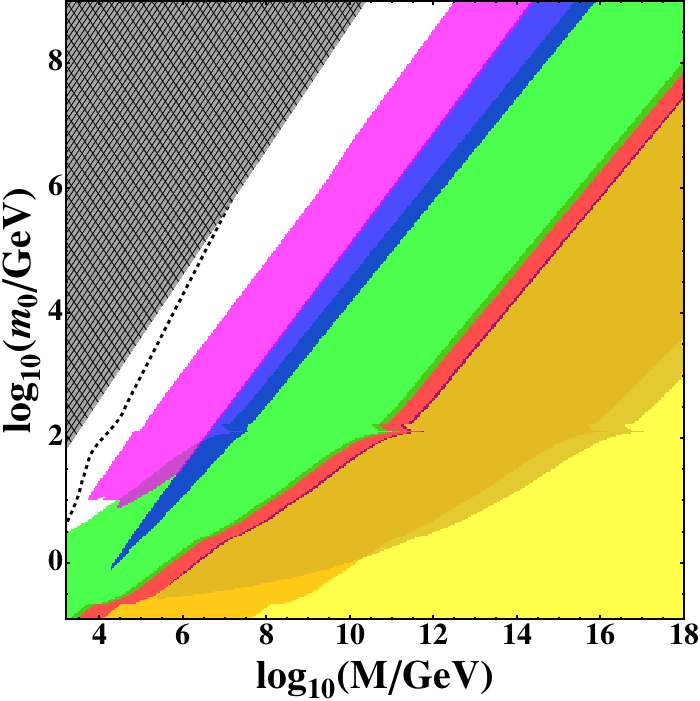}
\vspace{0.5cm}\\
	\includegraphics[width=0.25\textwidth]{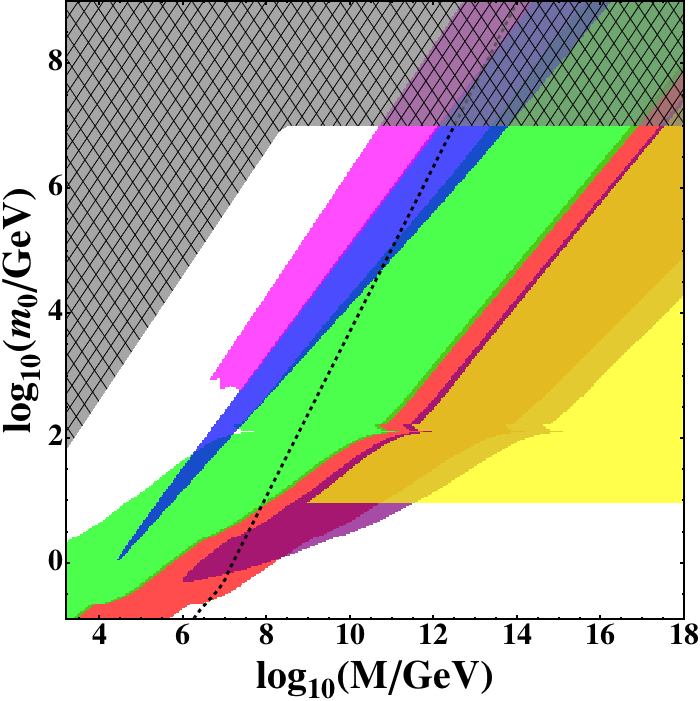}
	\includegraphics[width=0.25\textwidth]{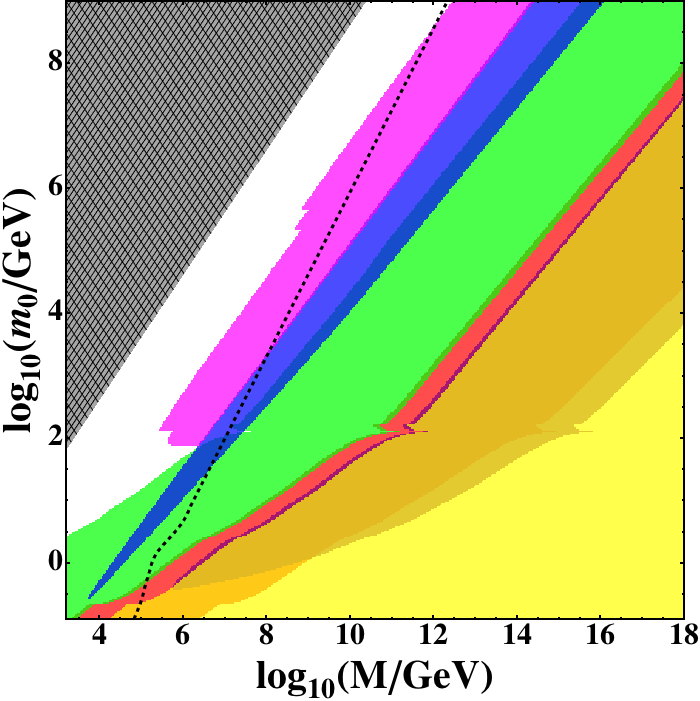}
	\includegraphics[width=0.25\textwidth]{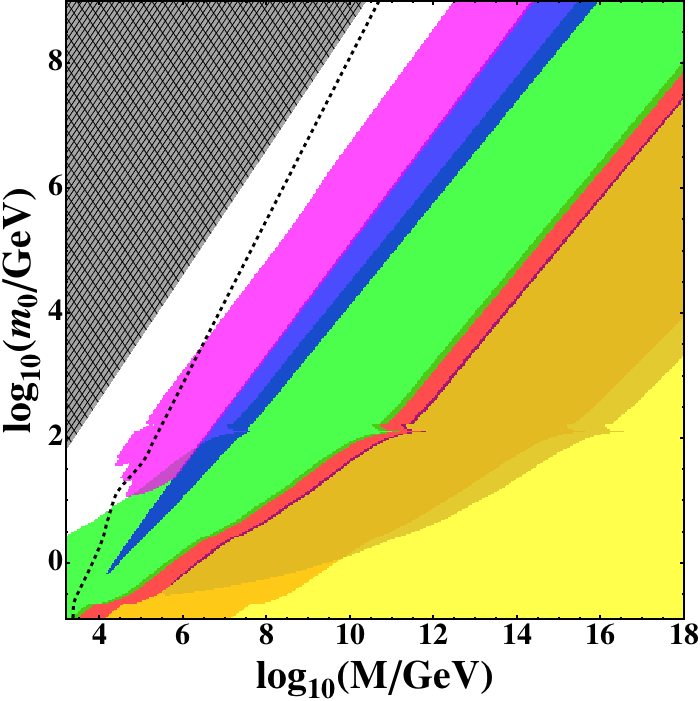}
	\vspace{-0.1cm}
	\caption{
Cosmological constraints on dark glueballs 
in the $M$--$m_0$ plane for decay scenario~\textbf{3}
with dominant dimension-6 operators and broken $C_x$.  
The upper two panels have $R=R_{min},\,R_{max}$, and the lower three panels have
fixed $R=10^{-9},\,10^{-6},\,10^{-3}$.  The black shaded region indicates where 
our theoretical assumptions fail, while to the left of the dashed line 
we find $R < R_{min}$. 
\label{fig:dec_s3}}
\end{figure}

The third decay scenario~\textbf{3} has both dimension-6 and dimension-8 operators
with $y_{eff} = 1$ and $\chi_i=\chi_Y = 1$, and broken $C_x$.
This leads to $\zpp$ decays dominated by the dimension-6 operator,
but decays of the $\opm$ only through the dimension-8 operators.  
As a result, the $\opm$ glueball is parametrically long-lived
relative to the $\zpp$ (and the other glueball states).  

We show the cosmological and astrophysical bounds on this scenario 
in Fig.~\ref{fig:dec_s3} for various values of the entropy ratio $R$.  
The upper two panels have $R=R_{min},\,R_{max}$ respectively,
and the lower three panels show $R=10^{-9},\,10^{-6},\,10^{-3}$.
As above, the grey shaded regions indicate where our theoretical
assumptions are not satisfied, and the diagonal dashed lines 
have $R< R_{min}$ to their left, except in the $R=R_{max}$ panel.
Here, thermalization is maintained all the way to confinement
(and beyond) to the left of the line.

 Decays of both the $\zpp$ and $\opm$ glueballs lead to relevant exclusions
in this scenario.  The $\zpp$ relic density tends to be much larger 
than the $\opm$ prior to decay, and produces the strongest constraints for small
values of $m_0/M$ when it is long-lived.  For very long lifetimes and small $R$, 
it can make up all the DM as before.  However, larger values
of $m_0/M$ lead to relatively short-lived $\zpp$ glueballs that
decay before the start of BBN.  In this case, the longer-lived
$\opm$ can decay late enough to disrupt nucleosynthesis or the CMB
in an unacceptable way.  Note as well that the region in which
the $\opm$ relic density is potentially large, it decays too quickly
to make up the dark matter.

\subsection{Decay Scenario 4: Dimension-6 Decays with Exact $C_x$}

\begin{figure}[ttt]
	\centering
${}$\hspace{0.8cm}
	\includegraphics[width=0.18\textwidth]{legend_no_decay}
\hspace{0.32cm}
	\includegraphics[width=0.25\textwidth]{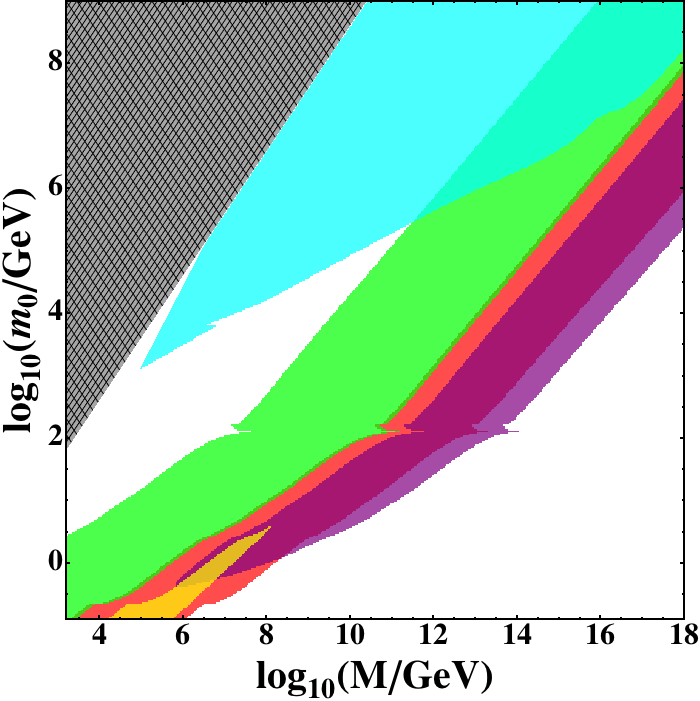}
	\includegraphics[width=0.25\textwidth]{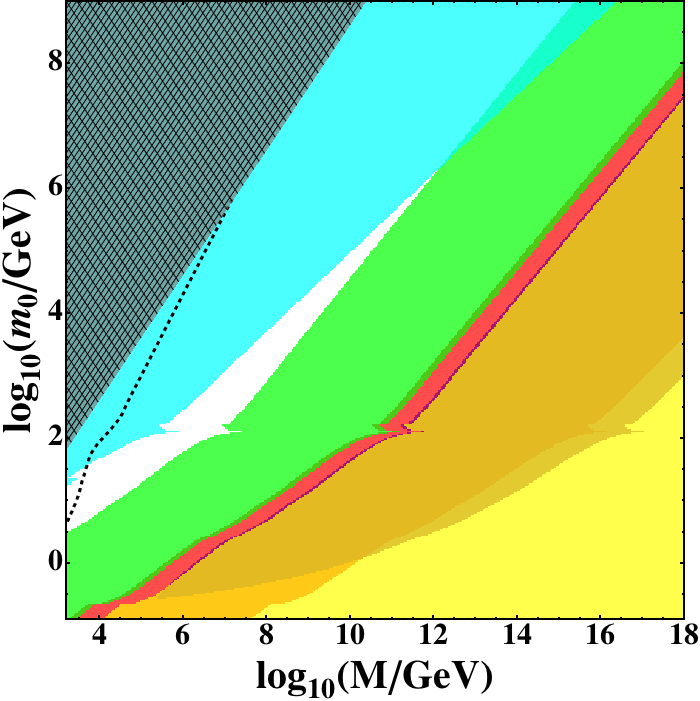}
\vspace{0.5cm}\\
	\includegraphics[width=0.25\textwidth]{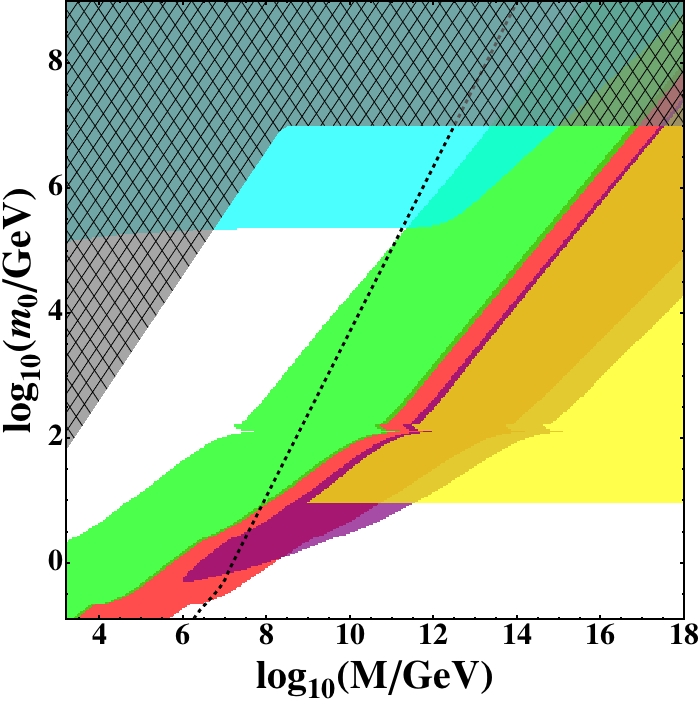}
	\includegraphics[width=0.25\textwidth]{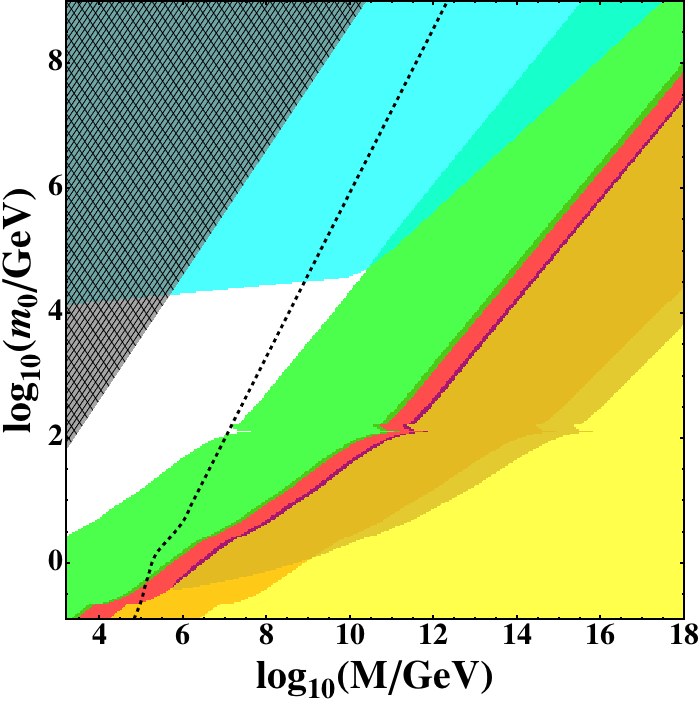}
	\includegraphics[width=0.25\textwidth]{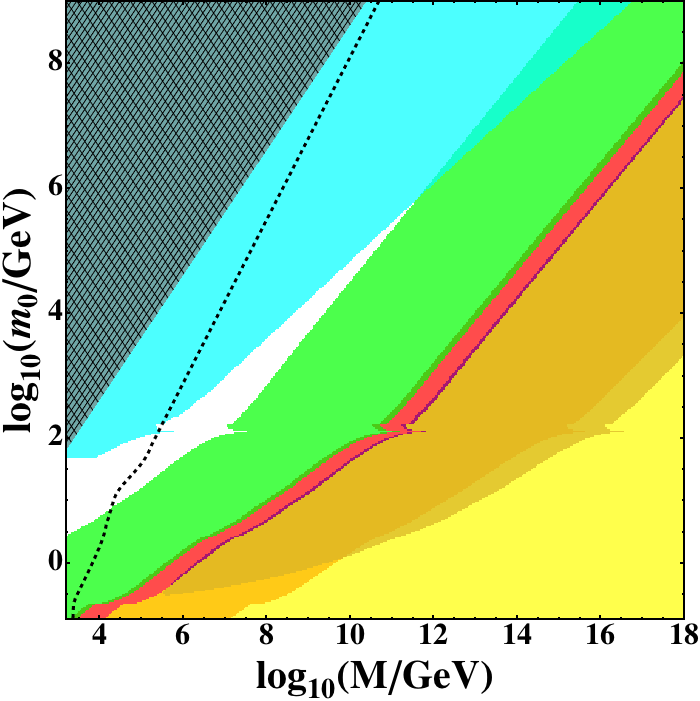}
	\vspace{-0.1cm}
	\caption{
Cosmological constraints on dark glueballs 
in the $M$--$m_0$ plane for decay scenario~\textbf{4}
with dominant dimension-8 operators and conserved $C_x$.  
The upper two panels have $R=R_{min},\,R_{max}$, and the lower three panels have
fixed $R=10^{-9},\,10^{-6},\,10^{-3}$.  The black shaded region indicates where 
our theoretical assumptions fail, while to the left of the dashed line 
we find $R < R_{min}$. 
\label{fig:dec_s4}}
\end{figure}

  Our final decay scenario~\textbf{4} has has both dimension-6 
and dimension-8 operators with $y_{eff} = 1$ and $\chi_i= 1$, 
together with conserved $C_x$ (and $\chi_Y=0$).  The $\zpp$ mode
decays as in the previous scenario, but now the $\opm$ is stable.

The cosmological bounds on this scenario are shown
in Fig.~\ref{fig:dec_s4} for various values of the entropy ratio $R$.  
The upper two panels have $R=R_{min},\,R_{max}$ respectively,
and the lower three panels show $R=10^{-9},\,10^{-6},\,10^{-3}$.
As above, the grey shaded regions indicate where our theoretical
assumptions are not satisfied, and the diagonal dashed lines 
have $R< R_{min}$ to their left, except in the $R=R_{max}$ panel.
Here, thermalization is maintained all the way to confinement
(and beyond) to the left of the line.

    The exclusions on this scenario from the $\zpp$ are identical to
those on scenario~\textbf{3}.  However, the constraints from the $\opm$
are now from its relic density rather than the effects of its decays
on BBN and the CMB.  This state can make up the dark matter for a range
of values of its mass and the entropy ratio $R$.  Compared to the 
analogous scenario~\textbf{2}, the relic density of the $\opm$ tends
to be larger here because it experiences less dilution from the 
more rapid decay of the $\zpp$.

\section{Conclusions\label{sec:conclusions}}
  
In this work we have investigated the cosmological constraints on non-Abelian
dark forces with connector operators to the SM.  We have focused on the minimal
realization of such a dark force in the form of a pure Yang-Mills theory.
In the early universe, the dark gluons of such theories confine to form a
set of dark glueballs.  Connector operators allow the transfer of energy
between the visible~(SM) and dark sectors, modify the freezeout dynamics
of the glueballs, and induce some or all of the dark glueballs to decay.  
Late decays of glueballs can modify the standard predictions for BBN, 
the CMB, and cosmic ray spectra, while very long-lived or stable 
glueballs must not produce too much dark matter.  Using these considerations, 
we have derived strong constraints on the existence of new non-Abelian 
dark forces.

A significant new feature of our work compared to previous studies~\cite{Faraggi:2000pv,Feng:2011ik,Boddy:2014yra,Boddy:2014qxa,Garcia:2015loa,Soni:2016gzf,Forestell:2016qhc,Halverson:2016nfq,Acharya:2017szw,Soni:2017nlm}
is the inclusion of the heavier $\opm$ glueball species.  
This state can be parametrically long-lived or stable relative to the 
other glueballs.  It freezes out in conjunction with the $\zpp$,
with the $\zpp$ density forming a massive thermal bath, leading to 
a rich array of freezeout and decay 
dynamics~\cite{Pappadopulo:2016pkp,Farina:2016llk}. 
In general, the (pre-decay) relic density of the 
$\opm$ mode is much smaller than the $\zpp$.  Even so, the $\opm$
can sometimes yield the strongest cosmological bounds due to its
longer lifetime. Specifically, the $\zpp$ could decay before impacting 
standard cosmological processes such as BBN, 
while the $\opm$ decays late enough to directly interfere. 
In some cases, the $\opm$ glueball could even make up the
observed DM density. 

   Our study also concentrated on the dark gauge group $G_x = SU(N\!=\!3)$
with a lightest $\zpp$ glueball mass above $m_0\geq 100\,\mev$.  The constraints
found here could also be generalized to other dark gauge groups and lower masses.
A very similar glueball spectrum is expected for $SU(N > 3)$~\cite{Teper:1998kw}, 
but the confining phase transition will be more strongly first-order
and its effect on glueball freezeout deserves further 
study~\cite{Garcia:2015loa,Lucini:2012wq}.
For $G_x=SU(2),\,SO(2N\!+\!1),\,Sp(2N)$ there are no $C_x$-odd 
glueballs~\cite{Juknevich:2009ji,Jaffe:1985qp}, 
but otherwise we expect the constraints based 
on the $\zpp$ glueballs to be applicable here.  In the case of $SO(2N\!>\!6)$,
the $C_x$-odd states are expected to be significantly heavier than the
$\zpp$, and thus the additional constraints on the lightest $C_x$-odd
mode would typically be weakened.  Our focus on $m_0> 100\,\mev$
was motivated by the ranges of masses considered in studies of the 
effects of late decays on BBN, the CMB, and gamma rays.  
Limits from the CMB can still be applied to much lower masses, but 
those from BBN and gamma ray production will be very different.
We leave a study of lower glueball masses to a future work.

\section*{Acknowledgements}

We thank
Sonia Bacca, Nikita Blinov, David Curtin, Anthony Francis, Richard Hill, 
Jonathan Kozaczuk, Robert Lasenby, Randy Lewis, John Ng,
Maxim Pospelov, Adam Ritz, Josh Ruderman, Richard Woloshyn, and Yue Zhang 
for helpful discussions.  
This work is supported by the Natural Sciences
and Engineering Research Council of Canada~(NSERC), with D.~M. and K.~S.
supported in part by Discovery Grants and L.~F. by a CGS~D scholarship.
K.S. acknowledges the kind hospitality of both the NYU Center for Cosmology 
and Particle Physics and the School of Natural Sciences at 
the Institute for Advanced Study, Princeton.
TRIUMF receives federal funding via a contribution agreement 
with the National Research Council of Canada.


\appendix

\section{Thermalization Rates\label{sec:appa}}

  The collision term appearing in Eq.~\eqref{eq:trans1} relevant
for thermalization corresponds to the process
$X+X\to \text{SM}+\text{SM}$, and is given by
\beq
\Delta\mathcal{C} &=& \langle\sigma\,v\ccdot \Delta E\rangle\,\widetilde{n}_x^2
\label{eq:coll1}\\
&=& \int\!d\Pi_1\int\!d\Pi_2\,f_1f_2\,W(s)\,\Delta E \ , \nnmb
\eeq
where $E_1$ and $E_2$ are the initial-state energies,
$\Delta E = (E_1+E_2)$ is to be evaluated in the comoving frame,
$d\Pi_i = g_i\,d^3p_i/(2\pi)^32E_i$, $f_i$ are the equilibrium
distribution functions at temperature $T$, and the \emph{scattering kernel}
is defined by~\cite{Gondolo:1990dk,Edsjo:1997bg}
\beq
W(s) &=& 4E_1E_2\sigma v\\
&=& \frac{\mathcal{S}}{g_1g_2}
\int\!\frac{d^3p_3}{(2\pi)^32E_3}\int\!\frac{d^3p_4}{(2\pi)^32E_4}
(2\pi)^4\delta^{(4)}(p_1+p_2-p_3-p_4)\,\sum_{\{int\}}|\mathcal{M}|^2 \ .\nnmb
\eeq
Here, $\mathcal{S}$ is the symmetry factor for identical particles, 
$g_i$ are the numbers of degrees of freedom of the initial-state particles,
the sum runs over all internal degrees of freedom, and $|\mathcal{M}|^2$
is the squared matrix element for the reaction.  Note that this quantity
is Lorentz invariant, and can therefore only depend on $s =(p_1+p_2)^2$.

  Following Refs.~\cite{Gondolo:1990dk,Edsjo:1997bg}, the expression
of Eq.~\eqref{eq:coll1} can be reduced to a single integral if we
approximate the distribution functions by Maxwell-Boltzmann forms,
$f_i= \exp(-E_i/T)$:
\beq
\Delta\mathcal{C} &=& \frac{g_1g_2T^2}{32\pi^4}\int_{(m_1\!+\!m_2)^2}^{\infty}\!ds\;
p_{12}\,\mathcal{F}(\sqrt{s}/T)\,W(s)
\nnmb\\
&&\label{eq:coll2}\\
&=& 
\frac{g_1g_2}{32\pi^4}\,T^5\,\int_{x_+}^{\infty}\!dx\;
\sqrt{\left(x^2-x_+^2\right)\left(x^2-x_-^2\right)}\,
\mathcal{F}(x)\,W(s=x^2T^2) \ ,
\nnmb
\eeq
where $\mathcal{F}(x) = \left(K_1(x)+\frac{x}{2}\left[
K_0(x)+K_2(x)\right]\right) = [2K_1(x)+xK_0(x)]$ and $x_\pm = (m_1\pm m_2)/T$.

\subsection{Cross Sections for Dimension-8 Operators}

The relevant operator has the general form
\beq
\mathcal{O}_8 = \frac{A}{M^4}\,X_{\alpha\beta}^aX^{a\,\alpha\beta}\,F_{\mu\nu}^CF^{C\,\mu\nu}
\ ,
\eeq
where $F_{\mu\nu}^C$ is a SM field strength.
This operator generates $XX\to AA$ transfer reactions for $T \gg m_A,\,m_0$, 
as well as $XA\to XA$ elastic scattering.  Concentrating on $XX\to AA$,
the corresponding matrix element for $(p_1,a)+(p_2,b)\to (p_3,C)+(p_4,D)$ is 
\beq
\mathcal{M} = 4A\frac{s^2}{M^4}\delta^{ab}\delta^{CD}
(\epsilon_1^*\cdot\epsilon_2^*)(\epsilon_3\cdot\epsilon_4) \ ,
\eeq
where $a,b,C,D$ are ``colours'' and $\epsilon_i$ are polarization vectors.
From this expression, we find (neglecting possible masses)
\beq
W(s) = \frac{1}{\pi}\lrf{\tilde{g}_A}{\tilde{g}_x}\,A^2\,\frac{s^4}{M^8} \ ,
\eeq
where $\tilde{g}_x$ and $\tilde{g}_A = 2(N_A^2-1)$
are the dark and visible numbers of degrees of freedom.
The energy-transfer collision term is then
\beq
\Delta\mathcal{C} 
&=& \frac{\tilde{g}_x\tilde{g}_A}{32\pi^5}A^2
\left[\int_0^\infty\!dx\;x^{10}\mathcal{F}(x)\right]\,\frac{T^{13}}{M^8} \ .
\eeq
The integral is dominated by $x =\sqrt{s}/T \sim 10$, 
corresponding to scattering at the high end of the thermal distribution.

  The coupling $A$ can be obtained by matching to our previous results 
for dark gluon connector operators.  While there are several operators 
that can contribute, we keep only the $S$ component corresponding 
to the operator listed above, which yields
\beq
A_i = \frac{\alpha_i\alpha_x}{120}\chi_i \ ,
\eeq
with $A_i = Y,2,3$ for each of the SM gauge factors.
For $\chi_i \to 1$, the gluon contribution dominates with 
$\tilde{g}_A = 2(N_c^2-1)$, and we focus on it exclusively.
Note that since we are considering $T \gtrsim m_0 \gtrsim 100\,\mev$
and the integration is dominated by $\sqrt{s} \sim 10T$, we should always
be safely above the QCD confinement scale.

\subsection{Cross Sections for Dimension-6 Operators}

The operator of interest is now
\beq
\mathcal{O}_6 = \frac{B}{M^2}\,|H|^2\,X_{\alpha\beta}^aX^{a\,\alpha\beta} \ ,
\eeq
with 
\beq
B = \frac{\alpha_xy_{eff}^2}{6\pi} \ .
\eeq
To treat scattering through this operator, we should distinguish
between temperatures above and below the electroweak phase transition at
$T_{EWPT} \simeq 100\,\gev$.  Above the transition,
all the SM states are massless and we can treat the Higgs field
as a complex scalar doublet.  Below the transition, 
we must account for masses.  

  For $T > T_{EWPT}$, the dominant transfer reaction is $X+X\to H+H^{\dagger}$, 
for which the scattering kernel is
\beq
W(s) = \frac{1}{\pi}\frac{1}{\tilde{g}_x}\,\frac{B^2}{M^4}s^2 \ .
\label{eq:kernH}
\eeq
This yields the collision term
\beq
\Delta\mathcal{C} = \frac{\tilde{g}_x}{32\pi^5}\,B^2\,
\left[\int_0^{\infty}\!dx\,x^6\mathcal{F}(x)\right]\,\frac{T^9}{M^4} \ ,
\label{eq:coll6a}
\eeq
where now the integral is dominated by $\sqrt{s} \sim 6T$.  

  Below the transition temperature, we have $f\bar{f}$, $hh$, $W^+W^-$, and $ZZ$
final states at leading order.  Their contributions to the scattering kernels are
\beq
W_f(s) &=& \frac{N_c^{(f)}}{\pi}\frac{1}{\tilde{g}_x}
\frac{B^2}{M^4}s^2\lrf{m_f^2}{s}
\lrf{s}{s-m_h^2}^2\left(1-\frac{4m_f^2}{s}\right)^{3/2} 
\label{eq:kernf}\\
W_h(s) &=& \frac{1}{4\pi}\frac{1}{\tilde{g}_x}\frac{B^2}{M^4}s^2
\left(1-\frac{4m_h^2}{s}\right)^{1/2}
\label{eq:kernh}\\
W_Z(s) &=& \frac{1}{4\pi}\frac{1}{\tilde{g}_x}\frac{B^2}{M^4}s^2
\lrf{s}{s-m_h^2}^2\left(1-\frac{2m_Z^2}{s}+\frac{12m_Z^4}{s^2}\right)
\left(1-\frac{4m_Z^2}{s}\right)^{1/2}
\label{eq:kernw}\\
W_W(s) &=& \frac{1}{2\pi}\frac{1}{\tilde{g}_x}\frac{B^2}{M^4}s^2
\lrf{s}{s-m_h^2}^2\left(1-\frac{2m_W^2}{s}+\frac{12m_W^4}{s^2}\right)
\left(1-\frac{4m_W^2}{s}\right)^{1/2}
\label{eq:kernz}
\eeq
These results only apply for $\sqrt{s} > 2m_i$; they are zero
otherwise. Note that for $\sqrt{s} \gg 2m_h,\,2m_f$, the sum
of these kernels is equal to the result of Eq.~\eqref{eq:kernH}.

\section{Cosmological Constraints \label{sec:appb}}

In this appendix we review the cosmological and astrophysical 
constraints imposed on massive dark glueballs.

\subsection{Decay Constraints from BBN}

  Particle decays during or after big bang nucleosynthesis~(BBN)
can modify the primordial abundances of light elements including
tritium, deuterium, helium, and lithium~\cite{Kawasaki:2004qu,Jedamzik:2006xz,Pospelov:2010hj,Cyburt:2015mya}.
The observed abundances of these light elements (with the exception of lithium)
agree well with the predictions of standard BBN when the baryon density
deduced from the CMB is used as an input~\cite{Cyburt:2015mya}.
If there was non-standard physics present during the era of BBN, 
such as the decays of dark glueballs to SM fields, the predictions 
the elemental abundance would be altered. Thus, constraints can be placed 
upon decays of glueballs after the onset of BBN.

Hadronic decays of a long-lived relic after $t \simeq 0.05\,\text{s}$
can modify the neutron~($n$) to proton~($p$) ratio and increase
the helium fraction through charge exchange reactions such as 
$\pi^-+p \rightarrow \pi^0 +n$,
or destroy light elements through spallation reactions like 
$n+ {^4\text{He}} \rightarrow \text{D} + p +2n$~\cite{Pospelov:2010hj,Cyburt:2015mya}. Electromagnetic decays are only constrained at later times,
after about $t\sim 10^{4}\,\text{s}$, since energetic electromagnetic
decay products emitted before this thermalize with the photon-electron plasma
before can they can destroy light elements by 
photodissociation~\cite{Kawasaki:1994sc,Pospelov:2010hj,Cyburt:2015mya}.

The combined effects of hadronic and electromagnetic decays on BBN 
have been studied in a number of works, 
including Refs.~\cite{Kawasaki:2004qu,Jedamzik:2006xz,Kawasaki:2017bqm}.
We apply the exclusions derived in Ref.~\cite{Kawasaki:2017bqm} to
place limits on decaying glueballs, using a simple interpolation to generalize
their results to arbitrary relic mass values between the range
$30\,\gev \leq m_x \leq 10^6\,\gev$ they studied, 
and matching to the appropriate set of final states.  
For masses outside these ranges, we apply the constraint for 
the nearest mass boundary.

\subsection{Decay Constraints from the CMB}

  Particle decays during or after recombination at 
$t\simeq 1.2\times 10^{13}\,\text{s}$
can modify the temperature and polarization spectra of the CMB.
They do so by injecting energy that increases the ionization fraction 
and temperature of the cosmological plasma.  In turn, this broadens the last 
scattering surface and alters the correlations among the temperature 
and polarization fluctuations~\cite{Chen:2003gz}.  

Detailed studies of the impact of such energy injection on the CMB
have been performed in Refs.~\cite{Padmanabhan:2005es,Zhang:2007zzh,Slatyer:2009yq,Finkbeiner:2011dx,Slatyer:2012yq,Cline:2013fm,Slatyer:2016qyl}.  
Corresponding limits on particle decays
based on the CMB measurements of Planck~\cite{Ade:2015xua} were extracted
in Refs.~\cite{Cline:2013fm,Slatyer:2016qyl}.  
Given the theoretical uncertainties in our calculation of the
pre-decay glueball yields, we apply a very simple parametrization
of the results of Ref.~\cite{Slatyer:2016qyl}: 
\beq
m_iY_i ~<~ (4.32\times 10^{-10}\,\gev)\lrf{\tau}{10^{24}\,\text{s}}\,
\mathcal{F}(\tau) \ ,
\label{eq:cmbbound}
\eeq
where $\mathcal{F}(\tau)$ accounts for the effects of early decays.
It is obtained by fitting to the curve of Fig.~4 of Ref.~\cite{Slatyer:2016qyl},
and is normalized to unity for $\tau \gg 1.2\times 10^{13}\,\text{s}$.
The form of Eq.~\eqref{eq:cmbbound} neglects mild dependences on
the mass of the decaying glueball and the specific final state, 
but these effects are smaller than the uncertainties in the calculation
of the pre-decay yield.  We also apply this limit to relic masses well
above the largest value of $m_x\sim 10\,\tev$ studied 
in Ref.~\cite{Slatyer:2016qyl} (and elsewhere).  
Such large masses lead to injections of highly energetic photons and
electrons that deposit their energy very efficiently in the cosmological
plasma~\cite{Slatyer:2009yq}.  As a result, we do not expect any major
loss of sensitivity for glueball masses well above $10\,\tev$.

Bounds on glueball decays can also be obtained from their effects
on the CMB frequency spectrum~\cite{Hu:1993gc,Feng:2003uy}.  
We find that these are subleading compared to those derived 
from BBN and the CMB power spectra.

\subsection{Decay Constraints from Gamma Rays}

  Glueballs with lifetimes greater than the age of the universe
$t_0 \simeq 4.3\times 10^{17}\,\text{s}$ can produce observable
signals in gamma ray and cosmic ray telescopes, even if their density is only
a small fraction of the total DM value.  Limits on the lifetimes
of decaying DM were derived in Ref.~\cite{Cohen:2016uyg} for dimension-6 
glueball decays and other final states over a broad range of masses
using galactic gamma ray data from \emph{Fermi}~\cite{Atwood:2009ez}.
With the theoretical uncertainty on glueball yields in mind, we 
use a simple parametrization of the limits on the glueball yield:
\beq
m_iY_i ~<~ (4.32\times 10^{-10}\,\gev)\lrf{\tau}{5\times 10^{27}\,\text{s}}\,
e^{t_0/\tau}\,e^{(10\,\gev/m_i)} \ ,
\eeq
where the last two factors account for the depletion of the signal if
the decay occurs before the present time and the loss of sensitivity
of \emph{Fermi} at lower masses~\cite{Massari:2015xea}.
This limit is fairly conservative and can be applied safely to
all dominant $\zpp$ decays, which occur on their own or shortly
after being created in a $\opm$ decay.

\bibliography{ref_inspire}

\end{document}